\input epsf.tex
\documentclass[12pt]{article}
\usepackage{graphicx}
\usepackage{amssymb,epsf}
\csname @addtoreset\endcsname{equation}{section}

\def\bseq{\begin{subequation}}  
\def\eseq{\end{subequation}}
\def\Bar#1{\overline{#1}}                       

\def\Tilde#1{\widetilde{#1}}                    

\newcommand{\beq}{\begin{equation}}
\newcommand{\eeq}{\end{equation}}
\newcommand{\bea}{\begin{eqnarray}}
\newcommand{\eea}{\end{eqnarray}}
\newcommand{\ena}{\end{eqnarray}}

\newcommand {\non}{\nonumber}

\renewcommand{\a}{\alpha}
\renewcommand{\b}{\beta}

\renewcommand{\d}{\delta}
\renewcommand{\th}{\theta}

\newcommand{\pa}{\partial}
\newcommand{\g}{\gamma}
\newcommand{\G}{\Gamma}

\newcommand{\e}{\epsilon}

\renewcommand{\l}{\lambda}
\renewcommand{\L}{\Lambda}

\newcommand{\Db}{\Bar{D}}
\newcommand{\Wb}{\Bar{W}}

\newcommand{\phib}{\bar{\phi}}

\newcommand{\thb}{\bar{\theta}}
\newcommand{\Phib}{\bar{\Phi}}

\newcommand{\Lb}{\bar{\Lambda}}
\newcommand{\Gbar}{\Bar{\Gamma}}
\newcommand{\Phibold}{\bold{\Phi}}
\newcommand{\Phibbold}{\bold{\bar{\Phi}}}

\newcommand{\ad}{{\dot{\alpha}}}
\newcommand{\bd}{{\dot{\beta}}}
\newcommand{\gd}{{\dot{\gamma}}}

\newcommand{\Del}{\nabla}
\newcommand{\Delb}{\overline{\nabla}}

\newcommand{\boldnabla}{  \nabla \hspace{-0.12in}{\nabla}}
\newcommand{\Wbar}{\overline{W}}

\newcommand{\intsup}{\int\!\! d^4xd^4\theta ~}     
\newcommand{\intch}{\int\!\! d^4xd^2\theta ~}      
\newcommand{\intach}{\int\!\! d^4xd^2\thb ~} 

\newcommand{\Fab}{\mathcal{F}^{\alpha\beta}}

\newcommand{\hb}{\overline{h}}
\newcommand{\se}{\mathcal{S}}
\newcommand{\ww}{\overline{W}^{\dot{\alpha}}\overline{W}_{\dot{\alpha}}}
\newcommand{\hm}{h_{12}}
\newcommand{\hbm}{\hb_{12}}
\newcommand{\hp}{(h_1\! +\! h_2)}

\begin{document}
\begin{titlepage}
\begin{flushright}
CPHT-RR023.0409\\
LPT-Orsay-09-25
\end{flushright}
\vspace{.3cm}
\begin{center}

{\Large \bf Nonanticommutative $U(1)$ SYM theories: Renormalization, 
fixed points and infrared stability}
\vfill
 
{\large \bf Marco S. Bianchi$^1$, Silvia Penati$^1$,~Alberto Romagnoni$^2$, \\ 
Massimo Siani$^1$}\\

\vspace{0.5cm}

{\small 
$^1$ Dipartimento di Fisica, Universit\`a di Milano--Bicocca and\\
INFN, Sezione di Milano--Bicocca, Piazza della
Scienza 3, I-20126 Milano, Italy\\

\vspace{0.1cm}
$^2$ Laboratoire de Physique Th�eorique, Univ. Paris-Sud and CNRS, F-91405 Orsay, and \\
CPhT, Ecole Polytechnique, CNRS, 91128 Palaiseau Cedex, France }\\

\end{center}
\vfill
\begin{center}
{\bf Abstract}
\end{center}
{Renormalizable nonanticommutative SYM theories with chiral matter in the adjoint 
representation of the gauge group have been recently constructed in [arXiv:0901.3094]. In the present
paper we focus on the $U_\ast(1)$ case with matter interacting through a cubic 
superpotential. For a single flavor, in a superspace setup and manifest background 
covariant approach  we perform the complete one--loop renormalization and compute the 
beta--functions for all couplings appearing in the action. 
We then generalize the calculation to the case of $SU(3)$ flavor matter with a cubic superpotential viewed 
as a nontrivial NAC  generalization of the ordinary abelian  $N=4$ SYM and its marginal deformations. 
We find that, as in the ordinary commutative case, the NAC $N=4$ theory is one--loop finite. We provide general 
arguments in support of all--loop finiteness. Instead, deforming the superpotential by marginal operators 
gives rise to beta--functions which are in general non--vanishing. We study the spectrum of fixed points and 
the RG flows. We find that nonanticommutativity always makes the fixed points unstable.}

\vspace{2mm} \vfill \hrule width 3.cm
\begin{flushleft}
e-mail: marco.bianchi@mib.infn.it\\
e-mail: silvia.penati@mib.infn.it\\
e-mail: alberto.romagnoni@cpht.polytechnique.fr\\
e-mail: massimo.siani@mib.infn.it
\end{flushleft}
\end{titlepage}

\section{Introduction}

Supersymmetric field theories can be defined on a nonanticommutative (NAC) superspace 
\cite{FL, KPT, DV, BGV, seiberg} where the spinorial variables satisfy 
$\{ \th^\a , \th^\b \} =  {\cal F}^{\a\b}$. The nontrivial spinorial algebra usually breaks supersymmetry 
down to $N = 1/2$.
The tensor ${\cal F}^{\a\b}$ has a stringy origin as the graviphoton field
which appears in the $N=2$ supergravity multiplet when taking the zero string length limit. 

NAC deformations of supersymmetric field theories have been extensively studied in four 
\cite{TY}-\cite{ferraraetal} and 
lower \cite{lowerdim} dimensions. In particular, since supersymmetry is partially broken a mandatory question 
is whether these theories maintain the robust renormalizability properties of their parent 
anticommutative theories. To this respect all NAC field theories investigated so far have two common features:
1) Nonanticommutativity is a mechanism of soft susy breaking;
2) Renormalizable NAC theories are not obtained from their ordinary parents by simply promoting  
products to NAC products in the original action but always require the addition of extra soft terms. 

One of the main issues to be addressed is the NAC formulation of gauge theories in interaction with 
chiral matter.
Recently, a renormalizable NAC deformation of SYM theories with matter in the adjoint 
representation of the gauge group has been proposed \cite{PRS}. This opens the possibility 
of investigating NAC deformations of SYM theories with extended supersymmetry. In particular, 
quantum consistent NAC deformations of $N=4$ SYM are now available which provide
the low energy dynamics of a stuck of D3--branes in the presence of a non--vanishing
RR two--form. This is an indispensable 
ingredient for generalizing the AdS/CFT correspondence to backgrounds with RR forms turned on in the
directions parallel to the branes.

As discussed in \cite{PRS} for the $SU({\cal N}) \otimes U(1)$ gauge group, adding adjoint matter to a NAC SYM theory with a non--trivial cubic superpotential 
leads to a theory which is not simply the natural generalization of the corresponding ordinary one obtained by promoting products to $\ast$--products in the classical action. In fact, the strict interplay between renormalizability and gauge invariance requires to assign a different coupling constant to the quadratic term for the abelian 
matter superfields in order to tune the renormalization of the abelian fields with the one for the non-abelians. 
This opens the possibility to add a renormalizable, $N=1/2$ and gauge invariant cubic superpotential. 
Moreover, it changes the gauge--matter coupling in vertices where abelian (anti)chirals 
are present.  As a crucial consequence, the evaluation of one--loop diagrams reveals that only $N=1/2$ susy and 
supergauge invariant divergent structures get produced. 
Therefore, a one--loop renormalizable action is obtained by 
adding all possible soft susy--breaking and supergauge invariant couplings allowed by dimensional analysis. 

Sufficient evidence for one--loop renormalizability has been given \cite{PRS}, but the complete renormalization has not been carried out yet.  
In fact, due to the non-trivial group structure, the form of the action is quite complicated
and the calculation of all one--loop divergent contributions would imply the evaluation of a large
number of diagrams. 

In order to avoid technical complications related to the group structure, in this paper we focus on the 
$U_\ast(1)$ case. The noncommutative $U_\ast(1)$ 
gauge theory is obtained from the non(anti)commutative $U({\cal N})$ theory in the limit ${\cal N} \to 1$. 
Despite the abelian nature of the generator algebra the resulting gauge theory is highly interacting
as a consequence of the non(anti)commutative nature of the $\ast$--product. 

In this case complications related to the different renormalization undergone by non--abelian and abelian 
superfields \cite{PRS} are absent and the general structure of SYM theories with matter in the adjoint
representation of the gauge group is rather simpler. 

We first consider the case of a single matter superfield  interacting with 
a cubic superpotential. We complete the one--loop renormalization of the theory and compute the corresponding 
beta--functions. 

We then generalize the calculation to the case 
of three adjoint chiral superfields in interaction through the superpotential
\bea
\label{superpot}
&& h_1 \intch \Phi_1 \ast \Phi_2 \ast \Phi_3 - h_2 \intch \Phi_1 \ast \Phi_3 \ast \Phi_2
\\
&~&~~~~~~~~~~~~~~~ \qquad 
+ \hb_1 \intach \Phib^1 \ast \Phib^2 \ast \Phib^3 - \hb_2 \intach \Phib^1 \ast \Phib^3 \ast \Phib^2 
\non
\eea
For  $h_1=h_2, \hb_1=\hb_2$ it exhibits a global $SU(3)$ invariance and can be interpreted as a nontrivial NAC deformation of the ordinary abelian $N=4$ SYM theory. Turning on nonanticommutativity breaks $N=4$ to $N= 1/2$.  
More generally, for  $h_1 \neq h_2$ and/or  $\hb_1 \neq \hb_2$ the $SU(3)$ symmetry is lost and the 
superpotential (\ref{superpot}) describes the NAC generalization of a marginally deformed 
\cite{LS,LM} $N=4$ SYM theory.  

We find that at one--loop the theory with  equal couplings is {\em finite} exactly like the ordinary $N=4$ counterpart. 
Using perturbative arguments based on dimensional considerations and symmetries of the theory we provide evidence that the theory should be finite at all loop orders. 
On the other hand, in the presence of marginal deformations UV divergences arise which in general prevent the 
theory from being at a fixed point. 

Both for the one and three--flavor cases we study the spectrum of 
fixed points and the RG flows in the parameter space. We find that nonanticommutativity always renders the fixed
points IR and UV unstable. Compared to the ordinary case, we loose the IR stability of the fixed point 
corresponding to the free theory ($h=\hb=0$ and $h_1=h_2, \hb_1 = \hb_2$). This is due to the fact that in the
NAC case the parameter space gets enlarged and new directions appear which drive the theories away from the 
fixed point.

The organization of the paper is the following: In Section 2 we define $U_\ast(1)$ NAC SYM theories with one and 
three chiral superfields in the adjoint representation of the gauge group, we discuss their 
gauge invariance and write their renormalizable actions according to the results of \cite{PRS}. In Section 3 we
present the one--loop renormalization for the case of a single matter field and the corresponding beta--functions. The same is done for the case of three chiral superfields in Section 4. Finally, in Section 5 we discuss the spectra
of fixed points and their stability. Conclusions follow plus an Appendix where all technical details required by the 
calculations are collected.

\section{$U_\ast(1)$ NAC SYM theories}

$N=(\frac12 , 0)$ NAC superspace is spanned by nonanticommutative coordinates 
$(x^{\a \ad}, \th^\a, \thb^\ad)$ satisfying
\beq
\{ \th^\a, \th^\b \} = 2 {\cal F}^{\a\b} \qquad \{ \thb^\ad, \thb^\bd \} = 0 \qquad [x^{\a\ad}, x^{\b \bd}] = 
[ x^{\a\ad}, \th^\b] = [ x^{\a\ad}, \thb^\bd] =0
\eeq
where ${\cal F}^{\alpha \beta}$ is a $2 \times 2$ symmetric, constant matrix. This algebra is
consistent only in euclidean signature where the chiral and antichiral sectors
are totally independent and not related by complex conjugation.

The class of smooth superfunctions on the NAC superspace is endowed with the 
NAC but associative product 
\beq
\phi \ast \psi ~\equiv~ \phi e^{- \overleftarrow{\pa}_\a {\cal F}^{\a \b}
\overrightarrow{\pa}_\b} \psi
~=~ \phi \psi - \phi \overleftarrow{\pa}_\a {\cal F}^{\a \b}
\overrightarrow{\pa}_\b \psi - \frac12 {\cal F}^2 \pa^2\phi \, {\pa}^2 \psi
\label{star} 
\eeq
where ${\cal F}^2 \equiv  {\cal F}^{\a \b} {\cal F}_{\a \b}$. 
(Anti)chiral superfields can be consistently defined by the constraints 
$\overline{D}_{\ad} \ast \phi = D_\a \ast \overline{\phi} =0$, where in chiral 
representation $D_{\a} = \pa_{\a} + i {\overline{\theta}}^{\ad} \pa_{\a \ad}$ and 
$\overline{D}_{\ad} =  \overline{\pa}_{\ad}$ (we use conventions of \cite{superspace}).

$U_\ast(1)$ supergauge group is defined as the limit of the NAC $U({\cal N})$ group when ${\cal N}=1$. Its 
elements are the chiral and antichiral superfields
\beq
g(x, \th, \thb) = e_\ast^{i \L(x, \th,\thb)}  \qquad , \qquad  \bar{g}(x, \th, \thb) = e_\ast^{i \Lb(x, \th,\thb)}
\eeq
which satisfy a noncommutative algebra. 
 
Given the non--abelian nature of $U_\ast(1)$ an adjoint representation can be defined according to the following 
prescription: A chiral superfield $\phi$ belongs to the adjoint representation of the gauge group if under supergauge transformations it transforms as
\beq
\phi \rightarrow \phi' = e_\ast^{i\L} \ast \phi \ast e_\ast^{-i\L}
\eeq
Equivalently, the transformation law for an antichiral superfield $\phib$ in the adjoint representation reads
\beq
\bar{\phi} \rightarrow \phib' = e_\ast^{i\Lb} \ast \phib \ast e_\ast^{-i\Lb}
\eeq
 
\vskip 15pt
As in the ordinary non--abelian case, supersymmetric $U_\ast(1)$ NAC Yang-Mills theories 
can be described in a manifestly covariant way by introducing a scalar prepotential $V$ 
in the adjoint representation of the gauge group transforming as
\beq
e_\ast^V \rightarrow  e_\ast^{V'} = e_\ast^{i \overline{\L}} \ast
e_\ast^V \ast e_\ast^{-i\L}
\eeq
Being the theory in euclidean signature, $V$ has to be {\em pure imaginary}, $V^\dag = -V$.

We define gauge covariant derivatives in superspace in the so--called
{\em gauge antichiral} representation \cite{superspace} as
\beq
{\nabla}_A \equiv (\nabla_\a , \overline{\nabla}_{\ad},
\overline{\nabla}_{\a \ad})
~=~ (D_\a ~,~ e_\ast^{V} \ast \overline{D}_{\ad} \, e_\ast^{-V} ~,~
-i \{ \nabla_\a, \overline{\nabla}_{\ad} \}_{\ast} )
\label{derivatives2}
\eeq
They act on superfields in the adjoint representation according to the prescription 
\beq
\label{derivatives1}
\Del \ast A \equiv  [ \Del , A ]_\ast = (D A)  - i [ \G , A]_\ast 
\eeq
where the connections are explicitly given by
\beq
{\G}_{\a}=0     \qquad , \qquad  \overline{\G}_{\ad} =
 ie_\ast^{V} \ast \overline{D}_{\ad} \, e_\ast^{-V}
\qquad , \qquad \overline{\G}_{\a \ad} = -i D_\a  \overline{\G}_{\ad}
\eeq
The corresponding field strengths are defined as $\ast$--commutators of supergauge
covariant derivatives
\beq
\overline{W}_{\ad} = -\frac12 [ \nabla^{\a},
\overline{\nabla}_{\a \ad} ]_\ast \qquad , \qquad 
\Tilde{W}_\a =
-\frac12 [ \overline{\nabla}^{\ad}, \overline{\nabla}_{\a \ad} ]_\ast
\label{W}
\eeq
and satisfy the Bianchi's identities $\nabla^\a \ast \Tilde{W}_\a + 
\overline{\nabla}^{\ad} \ast \overline{W}_{\ad}=0$.
In terms of gauge connections they are given by
\beq
\overline{W}_\ad = \frac{i}{2} D^\a \overline{\G}_{\a \ad} = D^2 \overline{\G}_\ad
\qquad , \qquad \Tilde{W}_\a = \frac{i}{2} \pa_\a^{\, \ad} \overline{\G}_\ad 
+ \frac{i}{2} [\overline{\nabla}^\ad , \overline{\G}_{\a \ad}]_\ast
\label{fieldstrengths}
\eeq
Covariantly (anti)chiral superfields can be defined according to 
$[\overline{\nabla}_\ad , \Phi]_\ast = 0$ and $[\nabla_\a , \overline{\Phi}]_\ast = 0$, 
respectively.

Specializing the results of \cite{PRS} to the $U_\ast(1)$ case the most general renormalizable action for a NAC SYM theory with one self--interacting chiral superfield in the adjoint representation of the gauge group
is given by (for simplicity we consider massless matter)  
\bea
\label{classaction1}
 S & = &\frac{1}{2g^{2}} \intach \Wb^\ad \ast \Wb_\ad  
 \non \\
 &~&~ + \intsup \Phi \ast \Phib  +  h\intch \Phi^{3}_\ast 
+  \hb\intach \Phib^{3}_\ast 
\non \\
&~&~  +i t_1 \Fab \intsup \thb^2~ \pa_\a^{\ \ad} \Gbar_{\b\ad} \ast \Phi \ast \Phib 
   + t_2 {\cal
  F}^2 \intsup \thb^2~ \Gbar^{\a\ad} \ast \Gbar_{\a\ad} \ast \Phib^3_\ast 
  \non \\ 
&~&~  +t_3 {\cal F}^2 \intsup \thb^2~ \Wb^\ad \ast \Wb_\ad \ast \Phi \ast \Phib 
\non \\
&~&~ + h_3 {\cal  F}^2 \intsup \thb^2~ \Phi \ast \Del^2 \Phi \ast \Del^2 \Phi 
  \non \\ 
 &~&~ +h_{4}
  {\cal F}^{2} \intsup \thb^2~ \Del^2 \Phi \ast \Phi \ast \Phib^2_\ast 
  +{h}_{5} {\cal F}^{2} \intsup \thb^2~ \Phi \ast \Phib^4_\ast 
 \eea
where $\Phi \equiv e^V_\ast  \ast \phi \ast e^{-V}_\ast$, $\Phib = \overline{\phi}$ 
are covariantly (anti)chiral superfields expressed in terms of ordinary (anti)chirals. We choose to indicate explicitly the $\ast$--product everywhere without distinguishing the cases where it actually coincides with the ordinary product. For example, it is easy to see that $\intach \Phib_\ast^3 = \intach \Phib^3$ up to superspace total derivatives.

We note that in contrast with the $SU({\cal N}) \otimes U(1)$  case \cite{GPR2} 
the pure gauge action contains only the NAC 
generalization of the standard quadratic term. In fact, it is easy to see that all the extra terms which need 
be taken into account in the  $SU({\cal N}) \otimes U(1)$ case for insuring renormalizability and gauge 
invariance are identically zero in the $U_\ast(1)$ limit.

More generally, we consider a theory with three different flavors in the (anti)fundamental representation 
of $SU(3)$, still interacting through a cubic 
superpotential. Again, using the results of \cite{PRS} the most general renormalizable action which respects
two global $U(1)$ symmetries is
\bea
\label{classaction2}
S & = &\frac{1}{2g^{2}} \intach \Wb^\ad \ast \Wb_\ad  + \intsup \Phi_i \ast \Phib^i 
  \non \\
&~&
\qquad \qquad \qquad +\intch \left( h_1 \Phi_1 \ast \Phi_2 \ast \Phi_3 - h_2 \Phi_1 \ast \Phi_3 \ast \Phi_2 \right)
\non \\
&~& \qquad \qquad \qquad + \intach \left( \hb_1 \Phib^1 \ast \Phib^2 \ast \Phib^3 - \hb_2  \Phib^1 \ast \Phib^3 \ast \Phib^2 \right)
\non \\ 
&&+i t_1 \Fab \intsup \thb^2~ \pa_\a^{\ \ad} \Gbar_{\b\ad} \ast \Phi_i \ast \Phib^i + t_2 {\cal
  F}^2 \intsup \thb^2~ \Gbar^{\a\ad} \ast \Gbar_{\a\ad} \ast \Phib^1 \ast \Phib^2 \ast \Phib^3 
  \non \\
&~&~ +t_3 {\cal F}^2 \intsup \thb^2~ \Wb^\ad \ast \Wb_\ad \ast \Phi_i \ast \Phib^i 
\non \\
&~&~ + \tilde
  h_3 \Fab \intsup \thb^2~ \Del_\a \Phi_1 \ast \Del_\b \Phi_2 \ast \Phi_3 + h_3
  {\cal F}^2 \intsup \thb^2~ \Phi_1 \ast \Del^2 \Phi_2 \ast \Del^2 \Phi_3 
\non \\ 
&~&~ +h_{4}^{(=)} {\cal F}^{2} \intsup \thb^2~ \sum_{i=1}^3 \Del^2 \Phi_i \ast \Phi_i \ast \Phib^i \ast \Phib^i 
\non \\
&~&~ +h_{4}^{(\neq)} {\cal F}^{2} \intsup \thb^2~ \sum_{i < j} \Del^2 \Phi_i \ast \Phi_j \ast \Phib^i \ast \Phib^j 
\non \\ 
&~&~ + {h}_{5} {\cal F}^{2} \intsup \thb^2~ \Phi_i \ast \Phib^i \ast \Phib^1 \ast \Phib^2 \ast \Phib^3 
\ena 
in terms of covariantly (anti)chiral superfields $\Phi_i$, $\Phib^i$. We note that one extra coupling $\tilde{h}_3$
is allowed in this case which would be trivially zero in the action (\ref{classaction1}), for symmetry reasons. 
The two global $U(1)$ charges for the matter superfields are $(1,-1,0)$ and $(0,1,-1)$ respectively, 
whereas antichiral superfields carry opposite charges.

The two actions are invariant under the following gauge transformations
\bea
\d \Phi_i = i [ \overline{\L} , \Phi_i]_\ast \qquad &,& \qquad
\d \Phib^i = i [ \overline{\L}, \Phib^i]_\ast
\nonumber \\
\d \overline{\G}_{\a\ad} =  [\overline{\nabla}_{\a\ad},  \overline{\L} ]_\ast \qquad &,& \qquad
\d \overline{W}_{\ad} = i [ \overline{\L} , \overline{W}_{\ad} ]_\ast
\label{Wtransf}
\eea
We note that except for the transformation of $\Gbar$ the right hand sides vanish when ${\mathcal F}^{\a\b} = 0$,
as it should in the ordinary $U(1)$ case (when taking the commutative limit matter in the adjoint representation of $U_\ast(1)$ is mapped into $U(1)$ singlets). 

In general, the cubic superpotential of (\ref{classaction2}) is a function of four independent couplings 
$h_1, h_2, \hb_1, \hb_2$. 
If we set $h_1 = h_2$ and $\hb_1=\hb_2$ the action (\ref{classaction2}) has a global $SU(3)$ invariance 
which can be thought of as related to the R--symmetry of an ordinary $N=4$ SYM theory. 
Therefore, we study the theory (\ref{classaction2}) as a non--trivial NAC deformation of the abelian $N=4$ 
SYM \footnote{At classical level, the NAC generalization of $N=4$ SYM theories has been studied in \cite{AI}
starting from an action which is the ordinary $N=4$ action with products promoted to $\ast$--products.}. We note that, while the ordinary $U(1)$ $N=4$ theory is a free theory of one vector superfield plus three 
chiral gauge singlets in the fundamental of $SU(3)$, the NAC deformation we propose is highly interacting.

More generally, if we set $h_1 = he^{i\pi \b}, h_2 = h e^{-i\pi \b}$ and 
$\hb_1 = \hb e^{-i\pi \bar{\b}}, \hb_2 = \hb e^{i\pi \bar{\b}}$ only the two global $U(1)$'s survive and we 
have the NAC generalization of beta--deformed theories \cite{LM}. We note that, being the theory in euclidean space with strictly real matter superfields, we need take the deformation parameters $\b, \bar{\b}$ to be pure 
imaginary in order to guarantee the reality of the action. In the ordinary anticommutative case 
supersymmetric theories with pure imaginary $\b$ have been studied in \cite{EMPSZ}.   

Both in the $N=4$ case and in its less supersymmetric marginal deformations, supersymmetry is broken to $N=1/2$ by the NAC superspace structure.  
  
\vskip 15pt
In order to perform perturbative calculations we use background field method \cite{superspace} 
suitably generalized to the NAC superspace \cite{GPR2}. 
As a result, at any given order in the loop 
expansion the contributions to the effective action are expressed directly in terms of covariant derivatives and
field strengths without any explicit dependence on the prepotential $V$. 

We split the Euclidean prepotential as $e^V _\ast
\rightarrow e^V_\ast \ast e^{U}_\ast$ where $U$ is the background prepotential and $V$ its 
quantum counterpart. 
Consequently, the covariant derivatives (\ref{derivatives2}) become
\beq
\nabla_\a = \boldnabla_\a =D_\a \qquad , \qquad \overline{\nabla}_{\ad} = 
e_{\ast}^V \ast \overline{\boldnabla}_{\ad}\ast e_{\ast}^{-V} = e_{\ast}^V \ast
(e_{\ast}^U\ast\Db_{\ad} \ e_{\ast}^{-U}) \ast e_{\ast}^{-V}
\label{bqsplitting}
\eeq
Covariantly (anti)chiral superfields in the adjoint representation are expressed in terms
of background covariantly (anti)chiral objects as 
\beq
\overline{\Phi} =
 \bold{\overline{\Phi}} \qquad \qquad , \qquad \qquad
   \Phi = e^V_{\ast} \ast \bold{\Phi}\ast e^{-V}_\ast
   =e_\ast^V \ast (e_\ast^U \ast \phi \ast e_\ast^{-U}) \ast e^{-V}_\ast
   \label{covchiral}
\eeq
We then split  $ \bold{\Phi} \to  \bold{\Phi} +  \bold{\Phi}_q$ and 
$ \bold{\Phib} \to  \bold{\Phib} +  \bold{\Phib}_q$, where  $\bold{\Phi}, \bold{\Phib}$ are
background fields and $\bold{\Phi}_q, \bold{\Phib}_q$ their quantum fluctuations. 

We break the invariance under quantum gauge transformations 
\cite{superspace,GPR2} by choosing gauge--fixing functions 
$f = \overline{\boldnabla}^2 \ast V$, $\overline{f} = {\boldnabla}^2 \ast V$, 
while preserving manifest invariance of the effective action and correlation
functions under background gauge transformations \cite{superspace,GPR2}.

The ghost action associated to the gauge--fixing is given in terms of background 
covariantly (anti)chiral FP and NK ghost superfields as
\beq
S_{gh} =  \int d^4x d^4 \theta ~ \Big[ \overline{c}' c - c'\overline{c} + .....+ \overline{b} b \Big]
\label{ghosts}
\eeq

In Ref. \cite{GPR2} the gauge--fixing procedure for NAC gauge theories has been discussed in detail. 
For the present scopes in the Appendix we summarize the procedure and collect the Feynman rules necessary 
for one--loop calculations.

We work in dimensional regularization, $n = 4 - 2\e$. All divergent integrals are expressed in terms of
the self--energy integral 
\bea
  {\cal S} =
\int d^n q ~\frac{1}{q^2(q-p)^2}
= \frac{1}{(4\pi)^2}~\frac{1}{\epsilon} + {\cal O}(1)
\label{selfenergy}
\eea

\section{One flavor case: Renormalization and $\b$--functions}

We first concentrate on the theory described by the action (\ref{classaction1}) and 
perform one--loop renormalization. 
 
Using Feynman rules listed in the Appendix we draw all possible one--loop divergent diagrams. 
A useful selection rule arises by looking at the overall power of the NAC parameter for a given diagram. 
In fact, as it is clear from 
the dimensional analysis of Refs. \cite{GPR2, PRS} divergent contributions can be at most quadratic 
in ${\cal F}^{\a\b}$. Since powers of ${\cal F}$ come from vertices and from the expansion of covariant 
propagators (see eqs. (\ref{vector}), (\ref{mixed})) it is easy to count the overall power of the 
NAC parameter and withdraw diagrams with too many ${\cal F}$'s.

  \begin{figure}
  \begin{center}
    \includegraphics[width=0.25\textwidth]{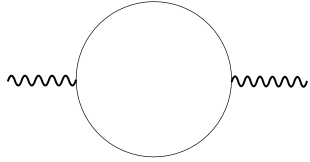}
   \end{center}
  \caption{Gauge self--energy diagram. }
  \label{Fig1}
\end{figure}

According to standard $D$--algebra arguments, in the NAC case as in the ordinary one divergent contributions 
to the gauge effective action come only from diagrams with a chiral matter/ghost quantum loop \cite{GPR2}. 
For the $U_\ast(1)$ theory the only potentially divergent contribution comes from the two--point diagram 
in Fig. 1 with interaction vertices arising from the expansion (\ref{vector}) of the covariant chiral propagator. 
Being the vertices of order
${\cal F}$ the result would be of order ${\cal F}^2$. Since dimensional analysis does not allow for self--energy
divergent contributions proportional to the NAC parameter we expect the divergent part of this diagram to 
vanish. In fact, by direct inspection it is easy to see that after $D$--algebra it reduces to a tadpole thus
giving a vanishing contribution in dimensional regularization. Therefore, the gauge action does not receive
any one--loop contributions. This is consistent with the result of \cite{GPR2} specialized to the ${\cal N}=1$
case.

We then concentrate on the renormalization of the gauge/matter part of the action (\ref{classaction1}).   
Using Feynman rules in the Appendix we select diagrams in Figs. 2, 3 as the only one--loop divergent diagrams. 
Diagrams $(2a,2c,2d,2e)$ are obtained from diagram $(3a)$ by expanding $1/\Box_{cov}$ as in (\ref{vector})
and writing $\Wbar \sim D \Gbar$.    
All internal lines are associated to ordinary $1/\Box$ propagators. 

By direct calculation it turns out that diagrams $(2d)$ and $(2e)$ cancel one against the other whereas 
the rest, after performing $D$--algebra, leads to the following one--loop effective action
\bea
\label{effective action1}
  \Gamma^{(1)}_{div} &=&
  \frac{1}{2g^{2}} \intach \bold{\Wb}^\ad \ast \bold{\Wb}_\ad  + \intsup \Phibold \ast \Phibbold\;
  \big[1+18h\hb\,\se\big] 
  \\ 
  && +h \intch \Phibold^{3}_\ast+ \hb\intach \Phibbold^{3}_\ast 
  \non \\ 
  && + i \, \Fab \intsup \thb^2~ \pa_\a^{\ \ad} \bold{\Gbar}_{\b \ad} \ast \Phibold \ast \Phibbold\; \big[t_1
  + 36\,( h\hb  - h \hb t_1)\,\se\big] 
  \non \\ 
  && + t_2 \, {\cal F}^2 \intsup \thb^2~ \bold{\Gbar}^{\a\ad} \ast \bold{\Gbar}_{\a\ad} \ast  \Phibbold^3_\ast 
  \non \\ 
  && + {\cal F}^2 \intsup \thb^2~ \bold{\Wb}^\ad \ast \bold{\Wb}_\ad \ast \Phibold \ast \Phibbold \; 
  \big[ t_3 + 36 \, (h \hb  - h \hb t_3  - h t_2) \,\se\big] 
  \non \\
  && + {\cal  F}^2 \intsup \thb^2~ \Phibold \ast (\boldnabla^2 \Phibold)^2_\ast  \; \left[ h_3 +
  (12\,g^2 h  -12\,g^2 t_1 h + 3\, g^2 t_1^2 h + 6\,  h h_4) \,\se\right] 
  \non \\
  && + {\cal F}^{2} \intsup \thb^2~ \boldnabla^2 \Phibold \ast \Phibold \ast \Phibbold^2_\ast \; \big[h_4 +
   (72\,  h \hb g^{2} t_1 -36\,h \hb g^{2} t_1^2   + 648\, h_3 h \hb^2  
  \non \\ 
  && \qquad\qquad\qquad\qquad\qquad\qquad \qquad\qquad\quad
  + 324\,h^2 \hb^{2} - 144\,h \hb h_4  + 36\,h h_5)\,\se\big] 
  \non \\ 
  && + {\cal F}^{2} \intsup \thb^2~ \Phibold \ast \Phibbold^4_\ast \;\big[ {h}_{5} + (108\, h \hb^{2}
  g^{2} t_1^2 + 216\, h \hb^{2} h_4 - 144\,h \hb h_5)\,\se\big]
  \non
\ena
where ${\cal S}$ is given in (\ref{selfenergy}).

Few comments are in order. First of all we note that the matter quadratic term does not receive 
gauge contributions. This is 
consistent with the results of Ref. \cite{PRS} where it was already shown that the abelian gauge  
quadratic term does not correct by terms proportional to $g^2$. The superpotential does not renormalize
thanks to 
the non--renormalization theorem which holds also in the NAC case. A similar behavior is exhibited by
the $t_2$--term which, at least at one--loop, seems to be protected from renormalization. However,
in this case we do not have any argument for expecting such a protection beyond one--loop. 

We now proceed to the renormalization of the theory by defining
renormalized coupling constants as 
\begin{eqnarray}
\label{ren}
&&\Phi = Z^{-\frac12} \Phi_B \qquad \qquad \qquad  \; \; \Phib = \bar{Z}^{-\frac12} \Phib_B
\\
&&g = \mu^{-\e} Z_g^{-1}  g_B \qquad \qquad \qquad \, h = \mu^{- \epsilon} Z_{h}^{-1} h_B \qquad \qquad \; \; \quad \bar{h} =
\mu^{- \epsilon} Z_{\bar{h}}^{-1} \bar{h}_B 
\nonumber \\
&&t_1 =  Z_{t_1}^{-1} t_{1\, B} \qquad \qquad  \: \: \: \:  \: \:  \, \quad  t_2 = \mu^{- \epsilon}
Z_{t_2}^{-1} t_{2 \, B} \qquad \qquad \; \; \;  t_3 =  Z_{t_3}^{-1} t_{3\, B}
\non \\
&&h_3 = \mu^{- \epsilon} Z_{h_3}^{-1} h_{3 \, B} \qquad \quad \quad \;
h_4 = \mu^{-2 \epsilon} Z_{h_4}^{-1} h_{4\,B} \qquad \qquad \; 
h_5 = \mu^{-3 \epsilon} Z_{h_5}^{-1} h_{5\,B} 
\non
\end{eqnarray}
where powers of the renormalization mass $\mu$ have been introduced in
order to deal with dimensionless renormalized couplings.
In order to cancel the divergences in (\ref{effective action1}) we set
\bea
\label{zeta}
  &&Z=\bar{Z}= 1 -
  18\,\frac{h\hb}{(4\pi)^{2}}\, \frac{1}{\epsilon} 
  \non \\ 
  &&Z_{\,h} h = h
  + 27\,\frac{h^2\hb}{(4\pi)^{2}}\,\frac{1}{\epsilon} \, \equiv \, h + \frac{h^{(1)}}{\e} 
  \non \\ 
  &&Z_{\,\bar{h}} \bar{h} = \hb
  + 27\,\frac{h\hb^2}{(4\pi)^{2}}\,\frac{1}{\epsilon} \, \equiv \, \hb + \frac{\hb^{(1)}}{\e} 
  \non \\ 
  &&Z_{\,h_3}\, h_3 =
  h_3 + \frac{27\, h \hb h_3 - 12\, g^2 h +12\,  g^2 h t_1 - 3\, g^2
  h t_1^2 - 6 \,h h_4}{(4\pi)^2 }\,\frac{1}{\epsilon} \, \equiv \, h_3 + \frac{h_3^{(1)}}{\e} 
  \non \\ 
  &&Z_{\,t_1}\,t_1 = t_1 +
  18\,(3\,t_1- 2)\,\frac{h\hb}{(4\pi)^{2}}\frac{1}{\epsilon}  \, \equiv \, t_1 + \frac{t_1^{(1)}}{\e} 
  \non \\
  &&Z_{\,t_2}\, t_2= t_2+\frac{ 27\,t_2
  h \hb}{(4\pi)^{2}}\, \frac{1}{\epsilon} \, \equiv \, t_2 + \frac{t_2^{(1)}}{\e} 
  \non \\ 
  &&Z_{\,t_3} \,t_3= t_3+\frac{54\,t_3 h \hb - 36\,h \hb +
  36\,h t_2}{(4\pi)^{2}}\, \frac{1}{\epsilon} \, \equiv \, t_3 + \frac{t_3^{(1)}}{\e} 
  \non \\
  &&Z_{\,h_4}\, h_4 = h_4
  + \frac{180\,h\,\hb\,h_4 - 36\,h {h}_{5} + 36 \,h \hb g^{2} t_1^2 -
  72\,  h \hb g^2 t_1 - 648\, h_3 h \hb^2 - 324\,
  (h \hb)^2}{(4\pi)^{2}}\,\frac{1}{\epsilon}  
  \non \\
  &~~&~~\qquad   \equiv \, h_4 + \frac{h_4^{(1)}}{\e} 
  \non \\ 
  &&Z_{\,h_5}\, h_5 = h_5
  - \frac{108 \,h \hb^{2} g^{2} t_1^{2} + 216 \,h \hb^{2} h_4 -
  189 \,h \hb h_5}{(4\pi)^{2}}\,\frac{1}{\epsilon}  \, \equiv \, h_5 + \frac{h_5^{(1)}}{\e} 
\ena
We have chosen to renormalize the chiral and the antichiral superfields in the same way, although this is 
not forced by any symmetry of the theory.
We note that divergences can be cancelled without renormalizing the NAC parameter ${\cal F}^{\a\b}$.
Therefore, the star product does not get deformed by quantum corrections. 

We compute the beta--functions according to the general prescription 
\beq
\label{beta}
\b_{\lambda_{j}} = - \epsilon~\alpha_j~\lambda_{j} -\alpha_j~\lambda_j^{(1)} + \sum_{i}
\left( \alpha_i~\lambda_{i} \frac{\pa \lambda_j^{(1)}}{\pa \lambda_{i}} \right)
\eeq
where $\l_j$ stands for any coupling of the theory and $\alpha_j$ is its naive dimension. 
Reading the single pole coefficients $\l_j^{(1)}$ in eq. (\ref{zeta}) we finally obtain 
\bea
\label{beta1}
 && \b_g = 0
 \non \\
  &&\beta_{\,h}=\frac{54 \,
  h^{2} \,\hb}{(4\pi)^{2}}
\non \\ 
  &&\beta_{\,\hb}=
  \frac{54 \,h\,\hb^{2}}{(4\pi)^{2}} \non \\
  && \beta_{\,h_3} = \frac{1}{(4\pi)^2}  \left(  54\, h \hb h_3 - 24\, g^2
  h + 24\, g^2 h t_1 - 6\, g^2 h t_1^2 -12 \,h h_4 \right) 
\non \\
  &&\beta_{\,t_1}= \frac{36}{(4 \pi)^{2}} \, (3\, t_1-2)\,h \hb
\non \\ 
  &&\beta_{\,t_2} =  \frac{54\,t_2 h \hb}{(4\pi)^{2}}
\non \\
  &&\beta_{\,t_3}=  \frac{1}{(4 \pi)^{2}}  \left(108\, t_3 h \hb - 72\, h \hb + 72\, h t_2 \right) 
\non \\ 
  &&\beta_{\,h_4}= \frac{1}{(4\pi)^{2}} \left(  72 \,h \hb g^2 t_1^2 - 144\,
  h \hb g^2 t_1 - 1296 \,h_3 h \hb^2 -648\, (h\hb)^2 + 360\,h \hb
  h_4 - 72\, h h_5 \right)
\non \\
  &&\beta_{\,h_5}= \frac{1}{(4 \pi)^{2}} \left( -216 \,h \hb^{2} g^{2} t_1^{2}
  - 432 \,h \hb^{2} h_4 + 378 \,h \hb h_5 \right) 
\ena

 \section{Three--flavor case: Renormalization and $\b$-functions}

In this Section we consider the case of the NAC $U_\ast(1)$ gauge theory in interaction with matter in the adjoint 
representation of the gauge group and in the fundamental representation of a flavor $SU(3)$ group. Its action is given in (\ref{classaction2}). We note that in the case $h_1=h_2, \hb_1 = \hb_2$, setting ${\cal F}^{\a\b} = 0$ turns off all the interactions and we are back to  the ordinary free U(1) $N=4$ SYM theory.
On the other hand, the noncommutative nature of the star product allows us to construct even in the "abelian"
case non--trivial interacting theories which can be studied as NAC deformations of $N=4$ SYM. 
More generally, we will consider $h_1 \neq h_2, \hb_1 \neq \hb_2$ in order to take into account also marginal deformations. 

We perform one--loop renormalization of the theory. Comparing to the case of a single chiral field, we note that 
the couplings are exactly of the same kind but dressed by flavor except for the extra coupling $\tilde{h}_3$ which 
in the previous case was trivially zero. Therefore, in order to evaluate divergent diagrams, it is sufficient 
to generalize the previous calculations to take into account non--trivial flavor factors and add possible 
new contributions arising from the contraction of a $\tilde{h}_3$ vertex with the rest. Since the 
$\tilde{h}_3$ vertex has the same structure of the vertex obtained when first order expanding the  
$\ast$--product in the superpotential (see vertices $(5f)$ and $(5h)$ in (\ref{mixed})), 
the topologies of divergent diagrams are still the ones in Fig. 2, 3. 

From a direct evaluation of all the contributions, for the one--loop divergent part of the effective action we find 
(in order to shorten the notation we define $\hm \equiv h_1 - h_2$ and $\hbm \equiv \hb_1 - \hb_2$)
\bea
\label{effective action2}
&&\Gamma^{(1)}_{div} = \frac{1}{2g^{2}} \intach \ww
  +\intch \Phi_{i}\Phib^{i}\;\big[1 + \hm \hbm \,\se\big] 
  \\ && +
  h_1 \intch \Phi_1 \Phi_2 \Phi_3 -
  h_2 \intch \Phi_1 \Phi_3 \Phi_2 
  \non \\
 && + \hb_1 \intach \Phib^1 \Phib^2 \Phib^3
 - \hb_2 \intch \Phib_1 \Phib_3 \Phib_2  
 \non \\
&& + \tilde{h}_3 \, \Fab \intsup \thb^2~ \Del_\a \Phi_1 \ast \Del_\b \Phi_2 \ast \Phi_3 
\non \\
  && + {\cal F}^2 \intsup \thb^2~ \Phi_1 \Del^2 \Phi_2 \Del^2 \Phi_3 \left[ h_3 +
  \left(12\, g^2 \hm - 6\, g^2 t_1 \hm   + 3 \,g^2
  t_1^2 \hm  + 3 \hm h_4^{(\neq)}\right) \,\se\right] 
  \non \\ &&
  + i\,\Fab \intsup \thb^2~ \pa_\a^{\ \ad} \Gbar_{\b \ad} \Phi_i \Phib^i \Big[ t_1
  + 2 \, \hm \hbm ( 1-t_1)\se \Big] 
  \non \\ 
  && + t_2 \,{\cal F}^2 \intsup
 \thb^2~ \Gbar^{\a\ad} \Gbar_{\a\ad} \Phib^1 \Phib^2 \Phib^3 
 \non \\
  && + {\cal F}^{2} \intsup \thb^2~ \Wb^\ad \Wb_\ad \Phi_i \Phib^i \Big[t_3 +
  2 \, \Big( \hm \hbm -  \hm \hbm t_3  - \hm t_2 \Big) \, \se\Big]
\non \\ 
  && + {\cal F}^{2} \intsup \thb^2~ \sum_{i} \Del^2 \Phi_i \Phi_i \Phib^i \Phib^i \left\{
  h_4^{(=)} + \left[  (h_3 + \tilde{h}_3) \hm \hbm^2 \right. \right.
  \non \\
  && \left. \left. \qquad \qquad \qquad  - 2 \,h_1 h_2 \hb_{12}^2  -
  2 \,\hm \hbm \,h_4^{(\neq)} + \hm h_5
  - \frac12 \left( \tilde{h}_3^2 + 2 \,\hp\tilde{h}_3 \right) \hbm^2 \right] \,\se \right\} 
  \non \\ 
  && + {\cal F}^{2} \intsup \thb^2~ \sum_{i<
  j} \Del^2 \Phi_i \Phi_j \Phib^i \Phib^j \Big\{h_4^{(\neq)} + \Big[ 
    8 \hm \hbm g^{2} t_1 -4 \hm \hbm g^{2} t_1^2  - 8 \hm \hbm h_4^{(=)} \non \\
  && 
  \qquad \qquad \qquad \qquad \qquad 
  + 2(h_1^2 + h_2^2) \hb_{12}^{2}  + 2(h_3 + \tilde{h}_3) \hm \hbm^2   
  +  \left( \tilde{h}_3^2 + 2 \hp\tilde{h}_3 \right) \hbm^2
 \non \\
 &&  \qquad \qquad \qquad \qquad \qquad  \qquad \qquad  \qquad \qquad \qquad \qquad \qquad
 - 4 \hm \hbm  h_4^{(\neq)} + 4 \hm h_5 \Big]\, \se \Big\}
\non \\
&& + {\cal F}^{2}
  \intsup \thb^2~ \Phi_i \Phib^i \Phib^1 \Phib^2 \Phib^3 \Big[ h_{5} +
  \Big( 4 \, \hm \hbm^2 g^{2} t_1^2 + 2\,\hm \hbm^2 h_4^{(=)}
   \non \\
&& \qquad\qquad\qquad\qquad\qquad\qquad\qquad \qquad \qquad \qquad \qquad 
+ 3\,\hm \hbm^2
  h_4^{(\neq)} - 6\, \hm \hbm h_5 \Big)\,\se\Big]
  \non
\ena
As in the previous case the gauge sector of the theory does not receive divergent contributions. 
Moreover, the quadratic matter action does not receive contributions from quantum gauge fields. 
 
Renormalization is still performed by using renormalized field functions and coupling constants as defined 
in (\ref{ren}). Choosing the same renormalization constants for the three (anti)chiral superfields,
in minimal subtraction scheme we set 
\bea
\label{zeta3}
  &&Z_{i} = \bar{Z}_{i} =
  1- \frac{\hm\hbm}{(4\pi)^{2}} \frac{1}{\epsilon} 
  \non \\ 
  &&Z_{\,h_1} =
  Z_{\,\hb_1} = Z_{\,h_2} = Z_{\,\hb_2} = Z_{\,\tilde{h}_3} = 1
  + \frac{3\,\hm \hbm}{2(4\pi)^{2}}\frac{1}{\epsilon} 
  \\ &&
  Z_{\,h_3} \,h_3 = h_3 + \frac{3\, h_3 \hm \hbm - 24\, g^2 \hm +
  12\, g^2 t_1 \hm 
  - 6\, g^2 t_1^2 \hm - 6\,\hm h_4^{(\neq)}}{2(4\pi)^{2}} \frac{1}{\epsilon} \non \\
  &&Z_{\,t_1}\,t_1 = t_1 + (3\,t_1 - 2) \, \frac{\hm \hbm}{(4\pi)^{2}}\frac{1}{\epsilon} \non \\
  &&Z_{\,t_2} \,t_2 = t_2 + \frac{3\, \hm \hbm t_2}{2(4\pi)^{2}}\frac{1}{\epsilon} \non \\
  &&Z_{\,t_3} \,t_3 = t_3 + \frac{3\, \hm \hbm t_3 - 2\, \hm \hbm +
  2 \, \hm t_2}{(4\,\pi)^{2}}\frac{1}{\epsilon} \non \\
  &&Z_{\,h_4^{(=)}}\, h_4^{(=)} = h_4^{(=)} + \frac{1}{(4\pi)^2 } \left[ 2\,\hm \hbm
  h_4^{(\neq)} - \hm h_{5} + 2\,\hm \hbm h_4^{(=)} \right.
  \non \\ 
  && ~~~~~~~\qquad ~~~~~~~ \qquad \left. +  2\,h_1\,h_2\,\hbm^2 - (h_3 + \tilde{h}_3) \hm \hbm^2
  + \frac12 \left( \tilde{h}_3^2 +2\,\hp \tilde{h}_3 \right) \hbm^2 \right] \frac{1}{\epsilon}
  \non \\ 
  &&Z_{\,h_4^{(\neq)}}\, h_4^{(\neq)}
  = h_4^{(\neq)} + \frac{1}{(4\pi)^2 } \left[ 4\,\hm \hbm g^{2} t_1^{2}\, - 8\,\hm \hbm
  g^{2} t_1 - 2(h_1^2 +h_2^2) \hbm^{2} + 8\,\hm \hbm h_4^{(=)} \right.
  \non \\ && ~~~ \left. \qquad \quad + 6\,\hm \hbm h_4^{(\neq)}  
  -4\,\hm h_{5} - 2(h_3 + \tilde{h}_3)\hm \hbm^2
  - \left( \tilde{h}_3^2 + 2 \, \hp\tilde{h}_3 \right) \hbm^2  \right] \frac{1}{\e} 
  \non \\
  && Z_{\,h_5} \,h_5 = h_5 - \frac{1}{(4\pi)^{2}} \left( 4\,\hm \hbm^{2} g^{2} t_1^{2} +
  2\,\hm \hbm^{2} h_4^{(=)}  + 3\,\hm \hbm^{2} h_4^{(\neq)}
  -\frac{17}{2}\,\hm \hbm \hb_{5} \right) \frac{1}{\epsilon}
  \non
\ena
Finally, applying the prescription (\ref{beta}) we find the beta--functions of the theory
\bea
\label{beta3}
&&\beta_g =0
\non \\
  &&\beta_{\,h_1}=\frac{3}{(4\pi)^{2}} h_1 \hm \hbm
 \qquad \qquad ~\beta_{\,\hb_1}=
  \frac{3}{(4\pi)^{2}} \hb_1 \hm \hbm
  \non \\
\non \\
  &&\beta_{\,h_2}=- \frac{3}{(4\pi)^{2}} h_2 \hm \hbm
 \qquad \qquad \beta_{\,\hb_2}=
   - \frac{3}{(4\pi)^{2}} \hb_2 \hm \hbm 
  \non \\
  &&\beta_{\tilde{h}_3}= \frac{3}{(4\pi)^{2}}  \,\hm \hbm \tilde{h}_3
  \non \\ 
  &&\beta_{h_3}=
   \frac{1}{(4\,\pi)^{2}}  \left( 3\,\hm \hbm h_3 - 24 g^2 \hm + 12 \,g^2 t_1 \hm -6\, g^2
  t_1^2 \hm - 6\,\hm h_4^{(\neq)} \right)
    \non \\
  &&\beta_{\,t_1}= \frac{2}{(4\pi)^{2}} (3\,t_1-2)\,\hm \hbm
  \non \\
  \non \\ 
  &&\beta_{\,t_2} =  \frac{3}{(4\pi)^{2}} \,\hm \hbm t_2
  \non \\ 
  &&\beta_{\,t_3}=  \frac{1}{(4\pi)^{2}} \left( 6\, \hm \hbm t_3 - 4\, \hm \hbm + 4 \,\hm
  t_2 \right)
  \non \\ 
  &&\beta_{\,h_4^{(=)}} =
   \frac{1}{(4\,\pi)^{2}} \left[ 4\,\hm \hbm h_4^{(\neq)} + 4\,\hm \hbm
  h_4^{(=)} - 2\,\hm h_5 + 4\,h_1 h_2 \hbm^2 \right.
  \non \\ 
  &&~~~~~~~~~\qquad \qquad ~~~~~~~ \left. - 2\, (h_3 + \tilde{h}_3) \hm \hbm^2
  +\left( \tilde{h}_3^2 +2\,\hp\tilde{h}_3 \right) \hbm^2 \right]
   \non \\ 
   &&\beta_{\,h_4^{(\neq)}}=
   \frac{1}{(4\,\pi)^{2}} \left[  8\, \hm \hbm g^{2} t_1^{2} -
  16\, \hm \hbm g^{2} t_1 + 16 \,\hm \hbm
  h_4^{(=)} \right. 
  \non \\ && \qquad \qquad \qquad \quad  ~~~~~~~~~~~~~~\left.
  + 12 \,\hm \hbm h_4^{(\neq)} - 8\,\hm h_5 -4\,(h_1^2 +
  h_2^2) \hbm^{2} \right.  
  \non \\ 
  && ~~~~~~~~~~~~~~ \qquad \qquad \qquad \quad \qquad \left. - 4\,
  (h_3 + \tilde{h}_3) \hm \hbm^2 - 2 \left( \tilde{h}_3^2 +2\,\hp \tilde{h}_3 \right) 
  \hbm^2 \right] 
 \non \\ 
 &&\beta_{\,h_5} =
  - \frac{1}{(4\,\pi)^{2}}  \left(8 \,\hm \hbm^{2} g^{2} t_1^{2} +
  4 \,\hm \hbm^{2} h_4^{(=)} + 6\,\hm \hbm^{2} h_4^{(\neq)} - 17 \,\hm \hbm
  h_5 \right)
  \non
 \ena

\begin{figure}
  \begin{center}
    \includegraphics[width=0.95\textwidth]{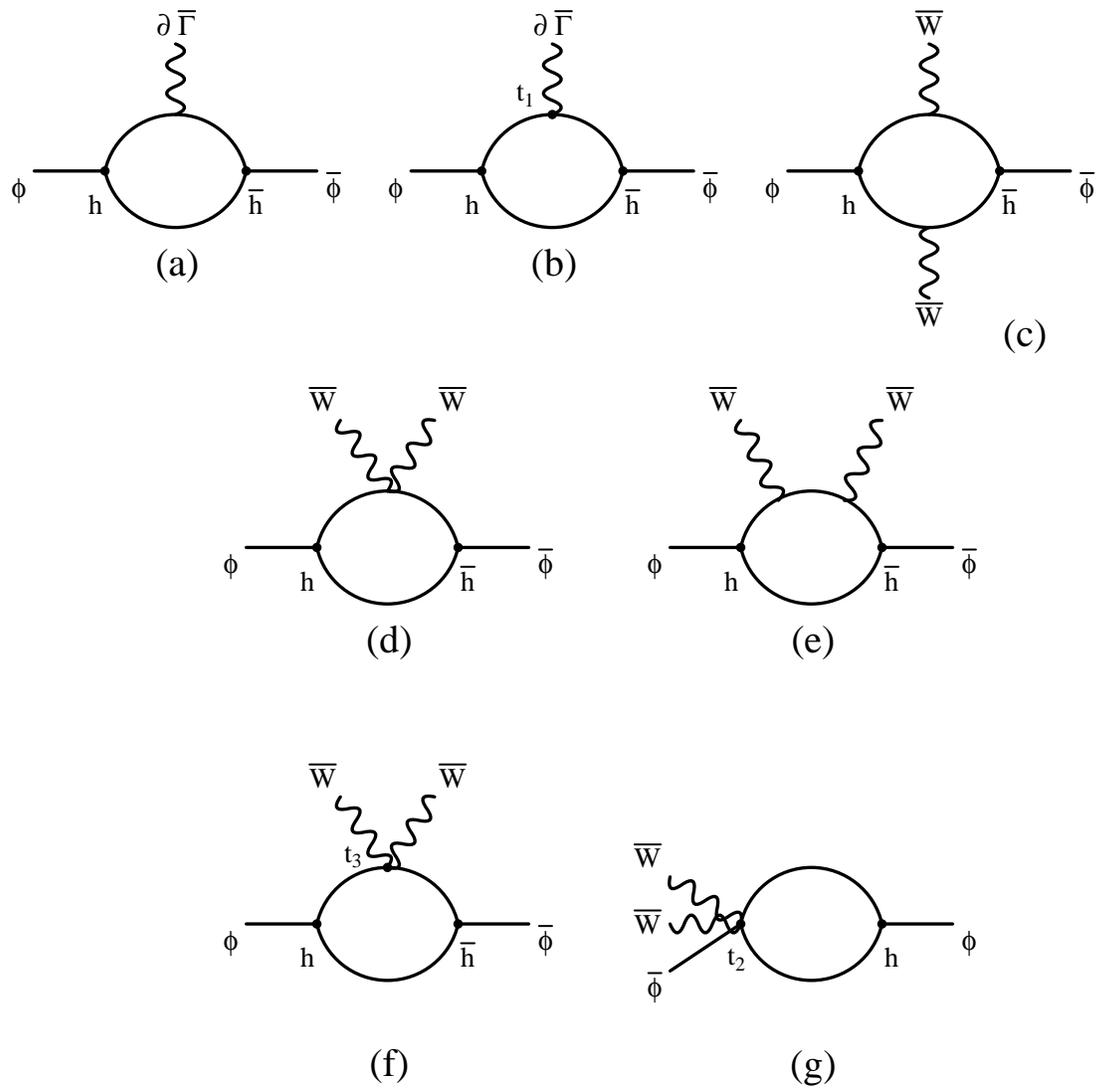}
   \end{center}
  \caption{One--loop diagrams contributing to the mixed sector. }
  \label{Fig2}
\end{figure}

\begin{figure}
  \begin{center}
    \includegraphics[width=0.95\textwidth]{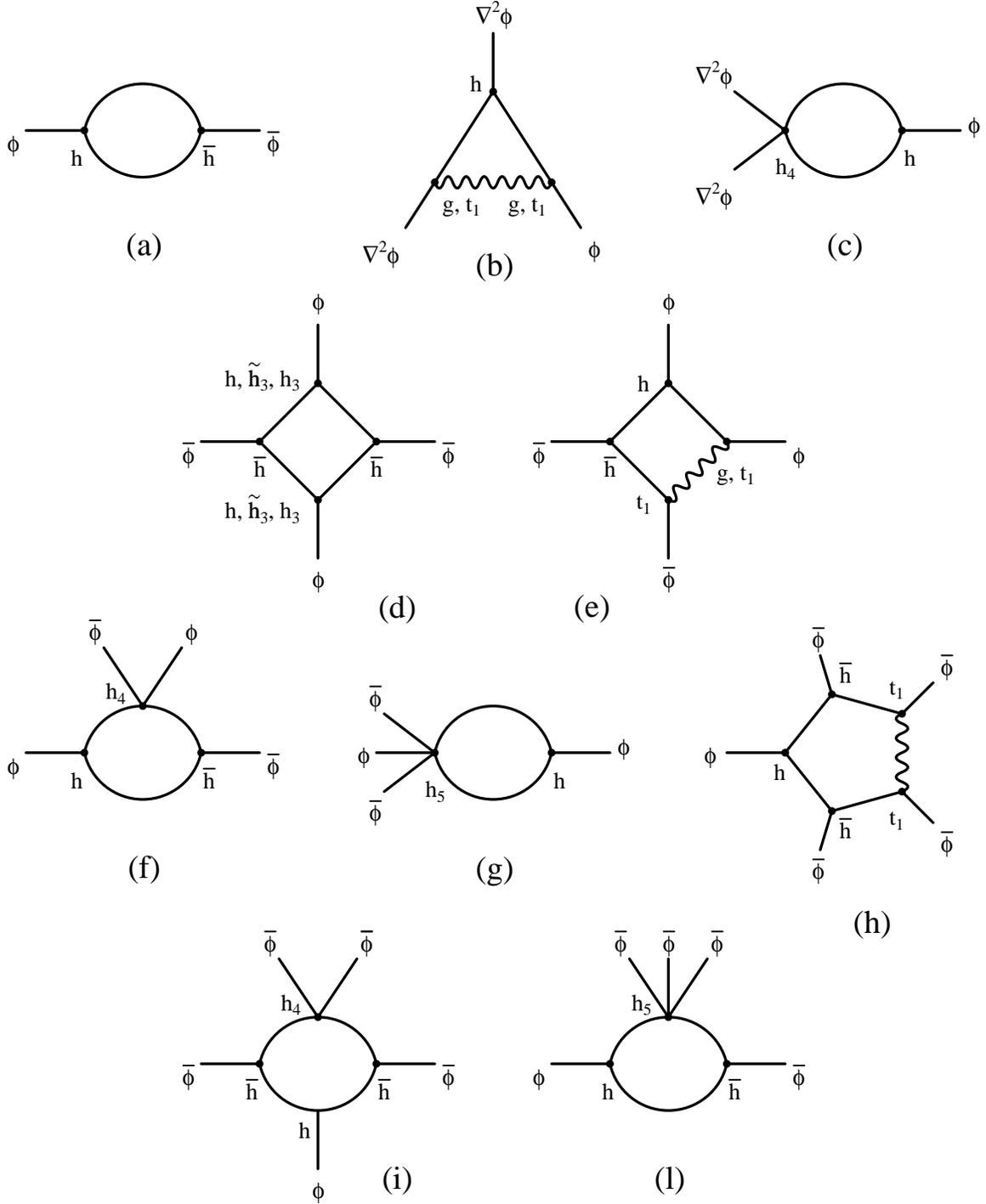}
   \end{center}
  \caption{One--loop diagrams contributing to the matter sector. }
  \label{Fig3}
\end{figure}

 \newpage

\section{Finiteness, fixed points and IR stability}

We now discuss the previous results for different choices of the
chiral couplings. We recall that we are working with euclidean
theories which are not subject to hermitian conjugation
constraints. In particular, $\Phi$ and $\Phib$ are independent real
superfields as well as the corresponding couplings $h$ and $\hb$.

We first consider the case of the theory with a single chiral
superfield. Referring to the results (\ref{effective action1}) we note
that all the divergences are proportional to (powers of) the
superpotential coupling $h$. Therefore, setting $h=0$ the theory turns
out to be one--loop finite and we have no need to add all possible
couplings in order to get a renormalizable theory. Precisely, the
following action
\beq
\label{h=0}
S = \frac{1}{2g^{2}} \intach \Wb^\ad \ast \Wb_\ad
+ \intsup \Phi \ast \Phib + \hb\intach \Phib^{3}_\ast
\eeq
is perfectly consistent at quantum level and one--loop finite. 
 
Conversely, if we set $\hb=0$ while keeping the chiral superpotential
on we have few divergent contributions surviving in (\ref{effective
action1}). Making the minimal choice of setting to zero all the extra
couplings which do not get renormalized we find that the following
action is one--loop renormalizable
\bea
\label{hb=0}
S & = &\frac{1}{2g^{2}} \intach \Wb^\ad \ast \Wb_\ad
+ \intsup \Phi \ast \Phib + h\intch \Phi^{3}_\ast
\non \\
&~&~~~~~~~~~ + h_3 {\cal
F}^2 \intsup \thb^2~ \Phi \ast \Del^2 \Phi \ast \Del^2 \Phi
\eea
but not finite.  This result is consistent with what has been
found \cite{BFR2, GPR1, BFR} for the NAC ungauged Wess-Zumino model.

The fact that the theory is finite when we turn off the superpotential
in the chiral sector while tolerating a superpotential for antichirals
but not viceversa is a manifestation of the asymmetry between the
chiral and the antichiral sectors induced by the star product.

We now discuss the spectrum of fixed points for the most general case where all the couplings
are turned on. As already
seen, the theory is one--loop finite when we set $h=0$, independently
of the value of the other couplings. Therefore, $h=0$ defines an eight
dimensional hypersurface of fixed points.

However, $h =0$ does not exhaust the spectrum of fixed points. In
fact, by a quick look at the beta functions in (\ref{beta1}) we can
easily see that taking $h \neq 0$ there is another hypersurface of
fixed points given by
\bea
&&\hb = h_5 = t_2 =0 
\non \\
&&2 h_4 + g^2 ( t_1- 2)^2 = 0 
\eea
In any case, from the requirement for $\b_h, \b_{\hb}$ to vanish we are
forced to set either $h$ or $\hb$ equal to zero. This is due to the
fact that, despite the non--trivial gauge/matter interaction, the
matter quadratic term does not get corrections from gauge quantum
fields.  As a consequence, we do not have non--trivial $h(g), \hb(g)$
functions which describe marginal flows as it happens in ordinary
non--abelian SYM theories.

We study the stability of fixed points and compare the present situation with the corresponding 
anticommutative case, that is an ordinary abelian SYM theory perturbed by a cubic superpotential 
\beq
h \intch \Phi^{3} + \hb \intach \Phib^3
\eeq
where hermiticity requires $\hb$ to the complex conjugate of $h$.  

In the ordinary case the theory is 
simply a free gauge theory plus a massless Wess--Zumino model. The corresponding
one--loop $\beta$--functions go like  $\b_h \sim |h|^2 h $ and $\b_{\hb} \sim |h|^2 \hb$. 
Therefore, the only fixed point of the theory is $h=\hb=0$. The RG trajectories are drawn in Fig. 4 
where only the first and third quadrants have to be considered ($h\hb= |h|^2 \geq0$).  Therefore, 
the origin corresponds to an IR stable fixed point. 

We now consider the NAC case described by the general action (\ref{classaction1}).  
The great number of coupling constants forbids plotting global RG
trajectories; however, we can study the IR behavior of the theory on
lower dimensional hypersurfaces by temporarily keeping a certain number of couplings
fixed. First of all, since $\b_g = 0$ we can sit on hypersurfaces
$g = \bar{g}$ where $\bar{g}$ is a small constant. Moreover, we can identify the flows associated to $\b_h$ and $\b_{\hb}$ as a closed subset of equations. 

The main difference compared to the ordinary case is that now $h$ and $\hb$ 
are two {\em real} independent couplings. This has two consequences: 
1) The spectrum of fixed points is now given by the two lines $h=0$ and $\hb=0$;  
2) Since the product $h\hb$ can be either positive or negative we need extend the study of RG trajectories to the whole $(h,\hb)$ plane. 
 
\begin{figure}
  \begin{center}
    \includegraphics[width=0.5\textwidth]{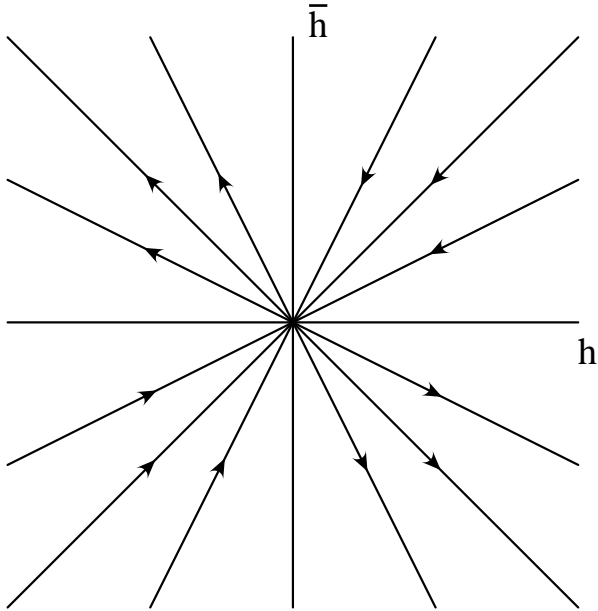}
   \end{center}
  \caption{Renormalization group trajectories near the $h=\hb=0$ fixed point.
  Arrows indicate the IR flows.}
  \label{fig:N=1}
\end{figure}

The configuration of RG trajectories is given in Fig. \ref{fig:N=1} where arrows indicate the IR flow. It is easy to see that the two axes $h=0$ and $\hb=0$  are lines of unstable fixed points.

In particular, we see that in this case the origin is neither 
an infrared nor an ultraviolet attractor. This is in contrast with the ordinary case where, as discussed above, the origin is an IR stable fixed point. The different behavior of the two theories can be traced back to the different 
hermiticity conditions which constrain the (anti)chiral coupling constants. 

Although the failure of the origin to be an IR attractor is conclusive, we can restrict the couplings to have the same sign (then studying the flows in the first and third quadrants) and investigate whether we can identify a region in the parameter space for which the origin is an infrared attractor. 

The $(h, \hb) = (0, 0)$ fixed point spans a seven dimensional hypersurface of fixed points corresponding 
to all possible values of the other couplings. We study RG trajectories on this hypersurface by linearizing in the rest of the couplings. 
   
The system of linearized equations we consider is $\mu d h_i/d \mu = \b_{h_i}, i = 3,4,5$, while the remaining 
equations decouple and have stability matrix with positive eigenvalues. Keeping $(h, \hb)$ slightly away from 
the fixed point, the eigenvalues of the stability matrix for the subset $(h_3,h_4,h_5)$ are approximatively 
\bea
  \rho_1 = -1.608\, h\, \hb \qquad \rho_2 = 232.788\,
  h\, \hb \qquad \rho_3 = 560.82\, h\, \hb
\eea
We see that the matrix vanishes at the fixed point but, as soon as we move away from the fixed point, there 
is at least one negative eigenvalue in any quadrant of the $(h,\hb)$--plane. The corresponding eigenvector 
represents an instability direction and leads to the conclusion that the origin is never an IR attractor whatever the range for $(h, \hb)$ is.

\vskip 15pt
We now consider the more interesting case of three flavors. As already
stressed, the theory (\ref{classaction2}) describes a NAC
generalization of the abelian $N=4$ SYM theory and theories obtained
from it by adding marginal deformations.

We remind that the ordinary abelian $N=4$ SYM theory is a free theory,
then necessarily finite. Marginal deformations can be added of the
form (we write them in a form which can be easily generalized to the
NAC case)
\bea
\intch \left( h_1 \Phi_1 \Phi_2 \Phi_3 - h_2  \Phi_1  \Phi_3 \Phi_2 \right)
+ \intach \left( \hb_1 \Phib^1 \Phib^2  \Phib^3 - \hb_2 \Phib^1  \Phib^3  \Phib^2 \right)
\non \\
\eea
which break supersymmetry down to $N=1$. In our notation $N=4$
supersymmetry is recovered for $h_1=h_2$ ($\hb_1 = \hb_2$ are the hermitian conjugates). 
The deformed theory is no longer finite since a divergent self--energy
contribution to the (anti)chirals appears at one--loop, proportional
to $h_{12} \hb_{12}$. It is easy to see that the free $N=4$ theory is a stable IR fixed point.

We now study what happens in the NAC case. Looking at the results
(\ref{effective action2}) the first important observation is that the
gauge beta--function is identically zero and all the other divergences
are proportional to powers of $\hm$ and $\hbm$. Therefore, setting
$h_1=h_2$ and $ \hb_1 = \hb_2$ kills all the divergences and the
theory is one--loop finite.  It follows that, at least at one--loop,
the NAC deformation does not affect the finiteness properties of the
$N=4$ SYM theory.

It is not difficult to provide general arguments for extending this
analysis to all loops.  First of all, the gauge sector cannot receive
loop corrections at any perturbative order. In fact, for dimensional
and symmetry reasons \cite{PRS} in the $U_\ast(1)$ case the only local
background structure which might be produced is the quadratic term
$\int \Wb^\ad \Wb_\ad$ with no powers of the NAC parameter in
front. As already discussed, any loop diagram that
we can draw contributing to the gauge sector is proportional to powers
of ${\cal F}^{\a\b}$ and then necessarily finite.
  
In the mixed and matter sectors, the constraints on the maximal power
of ${\cal F}^{\a\b}$ that we can have in divergent diagrams imply that
at least one chiral or one antichiral vertex from the superpotential
needs be present at order zero in the NAC parameter, therefore
carrying a coupling $h_{12}$ or $\hb_{12}$. Then, it is a matter of
fact that in the case of equal couplings all contributions vanish.

Therefore, on general grounds we conclude that the $U_\ast(1)$
deformation of the abelian $N=4$ SYM theory is all loop finite.

Exactly marginal deformations are obtained by adding marginal
operators to the action which do not affect the vanishing of the
beta--functions. In our case, taking $h_1 \neq h_2$ and/or
$\hb_1 \neq \hb_2$ means adding marginal operators. However, not all
of them turn out to be exactly marginal, at least at one--loop. In
fact, in order to have vanishing beta--functions away from the
symmetric point $\hm = \hbm = 0$ we need require either
\beq
\qquad ~h_1= h_2   \qquad \qquad  ( \tilde{h}_3 + 2 h_1 ) = 0 
\eeq
or 
\beq 
\hb_1 = \hb_2 \qquad \qquad t_2 = h_5 = 0 \qquad \qquad   g^2 (t_1^2 - 2t_1 + 4 ) + h_4^{(\neq)} = 0 
\eeq

In order to study the stability of the fixed points we can perform
an analysis similar to the previous one with the obvious substitutions
$h \rightarrow h_{12}$ and $\hb \rightarrow \hb_{12}$ plus the additional couplings 
which were not present in the one--flavor case. 

The flow equations for $h_{12}$ and $\hb_{12}$ still decouple from
the rest of the system and we can first study the IR behavior of the theory restricted to 
the $(h_{12}, \hb_{12})$ plane. With the suitable substitutions Fig. \ref{fig:N=1} is still valid
and provides two lines $h_{12}=0$ and $\hb_{12}=0$ of unstable fixed points. 

Restricting the range of $(h_{12}, \hb_{12})$ within the first and third quadrants and neglecting $t_2$ which 
has a trivial $\b$--function, we are left with a system of seven equations whose stability matrix can be studied 
in a neighborhood of the origin. The corresponding eigenvalues are approximatively
\bea
  \rho_1 = 3\, h_{12}\, \hb_{12} \qquad \rho_{2,3} = 6\,
  h_{12}\, \hb_{12} \qquad \rho_4 = -0.626\,
  h_{12}\, \hb_{12} \non \\ \rho_5 = 3.936\,
  h_{12}\, \hb_{12} \qquad \rho_6 = 11.674\,
  h_{12}\, \hb_{12} \qquad \rho_7 = 25.017\, h_{12}\, \hb_{12}
\eea
Again, the appearance of at least one negative eigenvalue for any choice of
the couplings leads to the conclusion that the $N=4$ theory is not an IR
attractor. This result is similar to what happens in the ordinary non--abelian SYM theories with
gauge group $SU(N \geq 3)$ \cite{nonabelian}, even if the two theories are not directly mappable one onto the other.

\section{Conclusions}

Deforming half of the Grassmannian part of the superspace could have bad consequences on the quantum 
behavior of field theories defined on it. In fact, due to the partial breaking of supersymmetry
the non--trivial cancellation between bosonic and fermionic divergences is not guaranteed anymore and 
we can expect a worsening in the UV behavior of the theories. Equivalently, the deformation introduces 
a ``bad'' dimensionful parameter which might induce the appearance of dimensionful 
momentum integrals with positive degree of superficial divergence. 
However, for all models investigated so far, a careful analysis has revealed that consistent 
completions of NAC deformations of ordinary theories can always be found for which renormalizability is  
preserved thanks to some global symmetries inherited from the parent theory.
 
In this paper we have continued on this line of investigation by performing the one--loop renormalization
of $U_\ast(1)$ SYM theories with 
matter in the adjoint representation of the gauge group, motivated by the idea of finding NAC generalizations
of ordinary SYM with extended supersymmetry. In general, the actions are not simply obtained 
from the ordinary ones by deforming the products, but contain suitable completions given in terms of 
all classical marginal operators which respect a given set of global symmetries.

We have first considered a SYM theory with a single chiral field self--interacting through a cubic superpotential.
Then, we have extended our analysis to the case of three matter fields interacting through a cubic superpotential which depends on four coupling constants, $h_1, h_2, \hb_1, \hb_2$. 
For  $h_1=h_2$ and $\hb_1=\hb_2$ the classical action exhibits a global $SU(3)$ symmetry and can be
interpreted as a NAC generalization of the ordinary $N=4$ SYM theory. More generally, for $h_1 \neq h_2$ 
and/or $\hb_1 \neq \hb_2$ it looks like the NAC generalization of marginal deformations of  $N=4$ SYM.

Since in the ordinary case $N=4$ SYM is finite, one of the questions we have addressed is whether 
finiteness survives in the NAC case. 
We note that, while in the ordinary $U(1)$ case finiteness is a trivial statement, being the theory free, 
its NAC generalization is highly interacting and the question becomes interesting. 
We have found that at one--loop the theory with  $h_1=h_2, \hb_1=\hb_2$ is indeed {\em finite}. Moreover,
based on general arguments we have provided a proof for the all--loop finiteness of the theory.    

More generally, we have considered theories in the presence of marginal deformations. In this case UV 
divergences arise which in general set the theory away from a fixed point. In the parameter space 
we have studied the spectrum of fixed points and the renormalization group flows. We have found that,
while in the ordinary $N=4$ case $h_1=h_2, \hb_1=\hb_2$ is an IR stable fixed point (free theory), in 
our case nonanticommutativity makes all the fixed points unstable. This is due to the fact that 
in the presence of extra marginal operators proportional to ${\cal F}^{\alpha \beta}$, the parameter
space gets enlarged and new lines of instability are allowed.
Even if our analysis is based on one--loop calculations, we have already enough information
for drawing qualitative conclusions on the effects that this kind of geometrical deformations have
on the RG flows: NAC theories resemble the non--abelian $SU(N\geq 3)$ ordinary theories for which
$N=4$ SYM is neither an IR nor an UV attractor. 

We focused only on massless theories but it is easy to convince that the addition of a mass 
term should not change the main features of the theories.

In order to simplify the analysis, we considered the $U_\ast(1)$ case. From the point of 
view of studying how renormalization works these theories are not too trivial. 
In fact, as already stressed, they are highly interacting. Therefore, the results obtained on
the finiteness in a subspace of the parameter space and, more generally, on the role of nonanticommutativity
on their UV and IR behavior are actually not {\em a priori} expected.

However, considering this example we have lost the non--trivial coupling between non--abelian 
and abelian superfields which is a peculiar feature of the NAC gauge theories. It would be then very 
interesting to consider the non--trivial $SU({\cal N}) \otimes U(1)$ case and investigate whether the 
obtained results survive.
In particular, it would be interesting to address the question of  finiteness. In fact, we expect that at one--loop the gauge sector would not receive divergent corrections since matter loops would cancel ghost loops, still giving
$\b^{(1)}_g =0$. In the matter sector new contributions proportional to $g^2$ would arise for the two and higher point functions. Therefore, as in the ordinary non abelian cases, we expect non--trivial surfaces of fixed points 
of the form $h_{12} = h_{12}(g), \hb_{12} = \hb_{12}(g)$.  The non--trivial question is whether this is
only a one--loop effect or it would arise as an actual feature of the whole quantum lagrangian. 
 
From a stringy point of view, our results are a further step towards a better understanding of  
the dynamics of D3-branes in the presence of non-vanishing RR forms and 
provide few hints for constructing gravity duals.

\vskip 25pt
\section*{Acknowledgements}
\noindent 

This work has been supported in part by INFN, PRIN prot.20075ATT78-002 and the European 
Commission RTN program MRTN--CT--2004--005104. The work of A.R. is supported by the European 
Commission Marie Curie Intra-European Fellowships under the contract N 041443.


\newpage
\appendix

\section{Background field method and Feynman rules} 

In this Appendix we collect all one--loop Feynman rules  obtained from the
actions (\ref{classaction1}, \ref{classaction2}) by applying the generalized background field method
developed in \cite{GPR2} for NAC super Yang--Mills theories with chiral
matter in a {\em real} representation of the gauge group.

\vskip 10 pt
\noindent
\underline{Gauge sector}

We work in gauge antichiral representation  \cite{superspace} for covariant derivatives and perform the 
quantum--background splitting according to
\beq
  \nabla_\a = {\boldnabla}_\a =D_\a \qquad , \qquad \overline{\nabla}_{\ad} =
  e_{\ast}^V \ast \overline{\boldnabla}_{\ad}\ast e_{\ast}^{-V} = e_{\ast}^V \ast
  e_{\ast}^U\ast\Db_{\ad} \ e_{\ast}^{-U} \ast
  e_{\ast}^{-V} 
\eeq
The derivatives transform covariantly with respect to quantum transformations
\bea
 \label{quantum}
&&
e_{\ast}^V \rightarrow e_{\ast}^{i \overline{\L}} \ast e_{\ast}^V \ast
e_{\ast}^{-i \L} \qquad , \qquad \qquad e_\ast ^U \rightarrow  e_\ast ^U
\nonumber \\
&& \nabla_A \rightarrow e_{\ast}^{i\overline{\L}} \ast \nabla_A \ast
e_{\ast}^{-i\overline{\L}}
\qquad , \qquad   {\boldnabla}_A \rightarrow  {\boldnabla}_A
\eea
with background covariantly (anti)chiral parameters, $\boldnabla_\a
\overline{\L} = \overline{\boldnabla}_{\ad} \L = 0$, and background  transformations
\bea
  \label{background}
 && e_{\ast}^V \rightarrow
e_{\ast}^{i \overline{\l}} \ast e_{\ast}^V \ast
e_{\ast}^{-i \overline{\l}} \qquad , \qquad \qquad
e_{\ast}^U \rightarrow e_{\ast}^{i \overline{\l}} \ast e_{\ast}^U \ast
e_{\ast}^{-i \l}
\nonumber \\
&& \nabla_A \rightarrow e_{\ast}^{i\overline{\l}} \ast \nabla_A \ast
e_{\ast}^{-i\overline{\l}} \qquad , \qquad   {\boldnabla}_A
\rightarrow  e_{\ast}^{i\overline{\l}} \ast {\boldnabla}_A \ast
e_{\ast}^{-i\overline{\l}}
\nonumber \\
\eea
with ordinary (anti)chiral parameters $\Db_\ad \l = D_\a \overline{\l}=0$.

The classical action 
\beq
 S_{gauge} = \frac{1}{2 g^2} \intach \overline{W}^{\ad} \overline{W}_{\ad}
\label{action1}
\eeq
for the gauge field strength defined in eq. (\ref{W}) 
is invariant under gauge transformations (\ref{quantum}) and (\ref{background}).
Background field quantization consists in performing gauge--fixing which explicitly
 breaks the (\ref{quantum}) gauge invariance while preserving manifest
 invariance of the effective action and correlation functions under
 (\ref{background}). Choosing as in the ordinary case the
 gauge--fixing functions as $f = \overline{\boldnabla}^2 \ast V$,
 $\overline{f} = {\boldnabla}^2 \ast V$ the resulting gauge--fixed
 action has exactly the same structure as in the ordinary
 case \cite{superspace} with products promoted to star
 products \cite{GPR2}. In Feynman gauge it reads
 \bea
&& S_{gauge} + S_{GF} + S_{gh}=
\non \\
&& \quad  -\frac{1}{2g^2} \intsup \Big[ e_\ast^{V} \ast
  {\overline{\boldnabla}}^\ad \ast e_\ast^{-V} \ast {D}^2
  (e_\ast^{V} \ast \overline{\boldnabla}_\ad \ast e_\ast^{-V}) ~+~
  V \ast (\overline{\boldnabla}^2 {D}^2 +
  {D}^2 \overline{\boldnabla}^2) \ast V \Big] 
  \non \\
&& \quad +  \intsup \Big[ \overline{c}' c - c'\overline{c} + .....
+ \overline{b} b \Big] 
  \label{gauge}
\eea
where ghosts are background covariantly (anti)chiral superfields and dots stand for higher order 
interaction terms. 
 
What we have reviewed so far holds for any NAC gauge theory, independently of the choice of
the gauge group. Now, we focus on the case we are interested in, that is $U_\ast(1)$ and determine the Feynman rules. 
   
Working out the quadratic part of the action from (\ref{gauge}) we find
\bea
  S + S_{GF}  \rightarrow &-& \frac{1}{2g^2} \intsup
  V \ast \hat \Box \ast V
\eea
where we have defined
\beq
  \hat{\Box} = \Box_{cov} -
  i\widetilde{\bold{W}}^\a \ast \boldnabla_\a - i \overline{\bold
  W}^\ad \ast \overline{\boldnabla}_\ad \qquad , \qquad \Box_{cov}
  = \frac12 \overline{\boldnabla}^{\a\ad} \ast \overline{\boldnabla}_{\a\ad}
\label{boxhat}
\eeq
We find convenient to rescale the gauge field as
\beq
V \rightarrow gV 
\label{rescale}
\eeq
Therefore, from the rescaled action we determine the covariant propagator
\beq \label{vectorprop} 
  \langle V(z)V(z') \rangle ~=~ \frac{1}{\hat{\Box}} \d^{(8)}(z - z')
\eeq
where $z \equiv (x^{\a\ad}, \th^\a, \thb^\ad)$. 

Expanding this expression in powers of the background fields it turns out that the covariant propagator
contains an infinite number of background--quantum interaction vertices. Precisely, we write
\beq
  \frac{1}{\hat{\Box}} \simeq \frac{1}{\Box_{cov}}
  + \frac{1}{\Box_{cov}} \ast \left( i\widetilde{\bold{W}}^\a \ast \boldnabla_\a + i \overline{\bold
  W}^\ad \ast \overline{\boldnabla}_\ad\right) \ast \frac{1}{\Box_{cov}}  + \cdots 
\eeq
and further expand $1/\Box_{cov}$. Since by direct inspection one can easily realize that terms proportional to 
$\widetilde{\bold{W}}^\a$ and $\overline{\bold W}^\ad$ never enter one--loop divergent diagrams,
we approximate 
\beq
\frac{1}{\hat{\Box}} \simeq \frac{1}{\Box_{cov}}
\eeq
and study in detail its expansion.

On a generic superfield in the adjoint representation of the gauge group we have
\bea \label{boxcov2}
  \Box_{cov} \ast \phi
  &\equiv& \frac{1}{2} [ \Delb^{\a\ad} ,[ \Delb_{\a\ad} , \phi]_\ast]_\ast  
  \non \\
  &=& \Box \phi -i [ \Gbar^{\a\ad}, \partial_{\a\ad} \phi]_\ast
  - \frac{i}{2} \left[
  (\partial^{\a\ad} \Gbar_{\a\ad}), \phi \right]_\ast
  - \frac{1}{2} \left[ \Gbar^{\a\ad}, \left[ \Gbar_{\a\ad}, \phi \right]_\ast \right]_\ast
\eea
where $\Box = \frac12 \pa^{\a\ad} \pa_{\a\ad}$ is the ordinary scalar kinetic term. 

Expanding the $\ast$--products and neglecting terms which never enter our calculations we find
\bea \label{boxcov3}
  \Box_{cov} = \Box + 2i \Fab
  (\pa_\a \Gbar^{\g\gd}) \pa_\b \pa_{\g\gd} -  {\cal F}^2
   (\pa^\a \Gbar^{\g \gd} ) \, (\pa^2 \Gbar_{\g\gd} ) \, \pa_\a
  + {\cal F}^2 
  ( \pa^\a \Gbar^{\g\gd} ) \, (\pa_\a \Gbar_{\g\gd}) \, \pa^2 + ...
\non \\
\eea
Inverting this expression we finally have 
\bea \label{Boxhat}
  \frac{1}{\Box_{cov}}
  &=& \frac{1}{\Box} \label{vector}
  \\
 &-&   \frac{1}{\Box} 2i~ \Fab (\pa_\a \Gbar^{\g\gd})
  ~ \pa_\b \pa_{\g\gd} \frac{1}{\Box} 
  - \frac{1}{\Box} 4~ \Fab
  (\pa_\a \Gbar^{\g\gd}) ~ \pa_\b \pa_{\g\gd} \frac{1}{\Box} ~ {\cal
  F}^{\eta\rho} ( \pa_\eta \Gbar^{\sigma\dot{\sigma}} )
  ~ \pa_\rho \pa_{\sigma\dot{\sigma}} \frac{1}{\Box} 
  \non \\ 
  &+& \frac{1}{\Box}  {\cal F}^2 \,
  (\pa^\a \Gbar^{\g \gd} ) \, (\pa^2 \Gbar_{\g\gd}
  ) \, \pa_\a \frac{1}{\Box} - \frac{1}{\Box} {\cal F}^2\, (\pa^\a \Gbar^{\g \gd} ) \, (\pa_\a \Gbar_{\g\gd}
  ) \, \pa^2 \frac{1}{\Box}+ \ldots
  \non
\eea
Here we recognize the ordinary
bare propagator $1/\Box$ plus a number of gauge interaction vertices. 
We note that all the interactions are proportional to the NAC parameter, as a peculiar feature of the 
$U_\ast(1)$ theory.

\vskip 15pt
\noindent
\underline{Matter sector}

In background field method we define  {\em   full} (anti)chiral superfields in the adjoint representation
of the gauge  group as
 \beq
 \label{covchiral2}
\overline{\Phi}
  = \bold{\overline{\Phi}} \quad \qquad  , \qquad \quad \Phi =
  e^V_{\ast} \ast \bold{\Phi}\ast e^{-V}_\ast =e_\ast^V \ast
  (e_\ast^U \ast \phi \ast e_\ast^{-U}) \ast e^{-V}_\ast
\eeq
where $ \bold{\Phi} \equiv e_\ast^U \ast \phi \ast e_\ast^{-U}$ and $\bold{\overline{\Phi}}$ are {\em background} covariantly (anti)chirals.

Under both quantum (\ref{quantum}) and background (\ref{background})
transformations the full (anti)chiral superfields 
transform covariantly  with parameters $\overline{\Lambda}$ and
$\overline{\lambda}$, respectively.

Under quantum transformations background covariantly (anti)chiral
fields transform as $ \bold{\Phi}' =
e_\ast^{i \Lambda}\ast \bold{\Phi} \ast e_\ast ^{-i \Lambda}$,
$ \bold{\overline{\Phi}}' = e_\ast^{i \overline{\Lambda}}\ast
\bold{\overline{\Phi}} \ast e_\ast ^{-i \overline{\Lambda}}$. Under background
transformations they both transform covariantly with parameter
$\overline{\lambda}$, $ \bold{\Phi}' = e_\ast^{i
  \overline{\lambda}}\ast \bold{\Phi} \ast e_\ast ^{-i
  \overline{\lambda}}$, $ \bold{\overline{\Phi}}' = e_\ast^{i
  \overline{\lambda}}\ast \bold{\overline{\Phi}} \ast e_\ast ^{-i
  \overline{\lambda}}$.  

Focusing the discussion on the $U_\ast(1)$ gauge group we now derive propagators and interaction vertices for matter in the actions (\ref{classaction1}, \ref{classaction2}) where we have
performed the rescaling (\ref{rescale}). 
Since one--loop divergent contributions are at most quadratic in the NAC
parameter, we list only Feynman rules entering these kinds of terms.

We split the actions (\ref{classaction1}, \ref{classaction2}) according to
\beq
S \equiv S_{gauge} + S_{matter} = S_{gauge} +  \intsup \Phib \ast \Phi + S_{int} 
\eeq
where $S_{gauge}$ is given in (\ref{gauge}) and $S_{int}$ is the rest of the matter actions
in (\ref{classaction1}, \ref{classaction2}) subtracted by the quadratic part. 

We concentrate on $S_{matter}$. Its quantization proceeds as usual. We first expand 
the full covariant quadratic action in terms of background covariantly
(anti)chiral fields (see (\ref{covchiral2})) 
\bea
  && \intsup ~ \Phibbold \ast e^{gV} \ast \Phibold \ast e^{-gV} \non \\
  &~&= \intsup \left\{ \Phibbold \Phibold + g \Phibbold
  [V, \Phibold]_\ast + \frac{g^2}{2} \Phibbold [V,
  [V,\Phibold]_\ast]_\ast + \ldots \right\} 
\label{exp2}
\eea
The first term in this expansion is
the kinetic term for background covariantly (anti)chiral fields.  In
particular, ghosts fall in this category so the same procedure
can be applied to the action (\ref{ghosts}), as well. The remaining terms give
rise to ordinary interactions with the quantum field $V$.

We perform the quantum-background splitting 
\beq 
\label{splitting}
\Phibold \rightarrow \Phibold + \Phibold_q \qquad ,\qquad  \Phibbold \rightarrow \Phibbold +
\Phibbold_q
\eeq
which allows to rewrite
\beq
S_{matter} = \intsup \Phibbold_q \Phibold_q + S'_{int}
\eeq
where $S'_{int}$ collects all the interaction vertices coming from $S_{int}$ after the splitting 
(\ref{splitting}) plus the extra interactions from (\ref{exp2}). 

Adding source terms
\beq
\intch j\Phibold_q + \intach \Phibbold_q \overline j 
\eeq 
and performing the gaussian integral in $\Phibold_q, \Phibbold_q$, the quantum partition function reads
\beq \label{effective}
  \mathbf Z[j, \overline j]  = \Delta_\ast \ast e^{S'_{int}( \frac{\delta}{\delta
    j}, \frac{\delta}{\delta \overline j})} \exp{\left[ - \frac12 \intsup
    \Big( j \ast \frac{1}{\Box_-} \ast \overline j  + \overline j \ast \frac{1}{\Box_+} \ast  j  \Big)
    \right]}
\eeq
where we have defined
\bea
&&\Box_+ = \Box_{cov} - i \widetilde{W}^\a \ast \nabla_\a
    -\frac{i}{2}(\nabla^\a \ast \widetilde{W}_\a)
\non \\
&&\Box_- = \Box_{cov} - i
    \Wbar^\ad \ast \Delb_\ad -\frac{i}{2}(\Delb^\ad \ast \Wbar_\ad)
\eea
and  $\Delta_\ast$ is the functional determinant
\beq
\label{eqn:Delta}
  \Delta_\ast = \int {\cal D} \Phibold_q {\cal D} \Phibbold_q
    ~ \exp{\intsup \Phibbold_q \Phibold_q}
\eeq

From the generating functional (\ref{effective}) we have two types of perturbative contributions, 
one from the expansion of $\Delta_\ast$ and one from the expansion of $\exp{(S'_{int})}$. 

As explained in \cite{superspace, GPR2}, $\Delta_\ast$ provides an additional, one--loop
contribution to the gauge effective action coming from matter/ghost loops.  
The corresponding Feynman
rules can be worked out by applying the ``doubling trick'' procedure \cite{superspace, GPR2}.
As a result, one--loop Feynman rules are obtained which can be formally read from the following 
effective action
\beq \label{effective-a}
\int d^4x d^4 \theta ~ {\rm Tr} \left\{ \overline{\xi}\Box \xi + \frac{1}{2} \left[ \overline{\xi}
D^2 (\overline{\boldnabla}^2 - \overline{D}^2)
\xi ~+~ \overline{\xi} ({\Box}_-- \Box ) \xi \right] \right\}
\label{effective-b}
\eeq
where $\xi ~,~ \overline{\xi}$ are {\em unconstrained} quantum fields with ordinary scalar propagator 
\beq
\label{prop}
\langle \xi (z) \bar{\xi}(z') \rangle = -\frac{1}{\Box} \d^{(8)} (z - z')
\eeq
and the first
vertex must appear once, and only once, in a one-loop diagram. 

The second type of contributions come from the expansion of $\exp{(S'_{int})}$ in (\ref{effective}).
The covariant matter propagators in this case are 
\bea
\langle \Phibold (z)\Phibbold (z')\rangle = - \frac{1}{\Box_-}\d^{(8)} (z - z')
 \non \\
\langle \Phibbold (z)\Phibold(z')\rangle = - \frac{1}{\Box_+}\d^{(8)} (z - z') 
\eea
which can be expanded according to 
\bea
\label{eqn:box_{cov}}
&&\frac{1}{\Box_-} \simeq \frac{1}{\Box_{cov}} + \frac{1}{\Box_{cov}} \ast \left( i \Wb^\ad \ast \Delb_\ad + \frac{i}{2} (\Delb^\ad \ast \Wb_\ad) \right) \ast \frac{1}{\Box_{cov}} + \cdots
\non \\
&&\frac{1}{\Box_+} \simeq \frac{1}{\Box_{cov}} + \frac{1}{\Box_{cov}} \ast \left( i \widetilde{W}^\a \ast \nabla_\a 
+ \frac{i}{2} (\nabla^\a \ast \widetilde{W}_\a) \right) \ast \frac{1}{\Box_{cov}} + \cdots
\eea
and contain an infinite number of interaction vertices between background gauge fields and quantum matter 
fields. As explained in the text, at one--loop divergent contributions arise only from the 
$\frac{1}{\Box_{cov}}$ part of the propagators. Therefore, we will set
\beq
\frac{1}{\Box_\pm} \simeq \frac{1}{\Box_{cov}}
\eeq
and further expand it as done in (\ref{vector}). 

Interaction vertices are obtained by working out the actual expression of $S'_{int}$ after the background--quantum 
splitting (\ref{splitting}). We list only the ones which effectively enter the evaluation of divergences. 
To keep the discussion more general we consider the three--flavor case. 
The one--flavor vertices are then easily obtained by dropping flavor indices and neglecting terms that, 
without flavors, vanish for symmetry reasons.  

We begin by considering the contributions (\ref{exp2}) coming from the quadratic action.
The only contributing vertex is $(5a)$ in Fig. 5
where $V$ is quantum and $\Phibold$ or $\Phibbold$ are background.
We then consider the $t_1, t_2, t_3$ interaction terms in (\ref{classaction2}). 
Because of the presence of a $\thb^2$ the $\ast$--products are actually ordinary products. 
The quantization proceeds by performing the splitting
(\ref{splitting}) on the (anti)chirals and expanding the connections and the field strength as follows
\bea \label{eqn:ti quant}
&&\Gbar_{\a\ad} \rightarrow - \Del_\a e^{-V} \overline{\boldnabla}_{\ad} e^{V} \rightarrow \Gbar_{\a\ad}
  - \Del_\a \left[ \overline{\boldnabla}_{\ad} , V \right]_\ast
  + \frac{1}{2} \Del_\a \left[ \left[\overline{\boldnabla}_{\ad} ,
  V \right]_{\ast} , V \right]_{\ast}
\non \\
&& \pa_{\beta\ad} \Gbar_\a^{~\ad} \, \longrightarrow \, \pa_{\b\ad} \Gbar_\a^{~\ad}
  - \pa_{\b\ad} \boldnabla_\a \left[ \overline{\boldnabla}^\ad,
  V \right]_\ast
\non \\
&& \Wbar_{\ad} \rightarrow -i \Del^2 e^{-V} \overline{\boldnabla}_{\ad} e^{V} \rightarrow \Wbar_{\ad}
  - i \Del^2 \left[ \overline{\boldnabla}_{\ad} , V \right]_\ast
  + \frac{i}{2} \Del^2 \left[ \left[\overline{\boldnabla}_{\ad} ,
  V \right]_{\ast} , V \right]_{\ast}
\eea
Collecting only the contributions which may contribute at one--loop we obtain vertices $(5b, 5d)$ 
where gauge is only background and vertex $(5c)$ where  $\Phibold$ or $\Phibbold$ are background.
We note that they all exhibit a gauge--invariant background dependence. 
We then turn to the pure matter interaction terms.
By splitting (anti)chiral superfields we find vertices $(5f - 5m)$. 

Collecting all the results, the explicit expressions for the vertices are 
\bea \label{mixed}
(5a) &\qquad& -2ig \, \thb^\ad \Fab  V (\pa_\a \Phi_i) \pa_{\b\ad} \Phib^i  
\non \\ 
\non \\
(5b) &\qquad& i t_1 \, \thb^2 \Fab (\pa_\a^{\ \ad} \bar{\G}_{\b \ad} ) \Phi_i \Phib^i
\non \\ 
\non \\ 
(5c) &\qquad& - i g t_1 \, \thb^2 \Fab ( \pa_{\a\ad} D_\b \bar{D}^\ad V) \Phi_i \Phib^i 
\non \\
\non \\ 
(5d) &\qquad& t_2 \, \thb^2 {\cal F}^2 \bar{\G}^{\a \ad}\bar{\G}_{\a \ad} \Phib_1 \Phib_2 \Phib_3
\non \\
\non \\ 
(5e) &\qquad& t_3 \, \thb^2 {\cal F}^2 \Wbar^{\ad} \Wbar_{\ad} \Phi_i \Phib^i
\non \\
\non \\ 
(5f) &\qquad& h_{12} \, \Phi_1 \Phi_2 \Phi_3 - (h_1 + h_2) \, \Fab \pa_\a \Phi_1 \pa_\b \Phi_2 \Phi_3  
- \frac12 h_{12} \, {\cal F}^2 \pa^2 \Phi_1 \pa^2 \Phi_2 \Phi_3
\non \\        
\non \\ 
(5g) &\qquad& \hb_{12} \, \Phib_1 \Phib_2 \Phib_3 - (\hb_1 + \hb_2) \, \Fab \pa_\a \Phib_1 \pa_\b \Phib_2 \Phib_3  
\non \\
\non \\ 
(5h) &\qquad&  \tilde{h}_3 \, \thb^2 \Fab \nabla_\a \Phi_1 \nabla_\b \Phi_2 \Phi_3  
+ \tilde{h}_3 \, \thb^2 {\cal F}^2 \nabla^2 \Phi_1 \nabla^2 \Phi_2 \Phi_3
\non \\
\non \\ 
(5i) &\qquad& h_3 \, \thb^2 {\cal F}^2 \nabla^2 \Phi_1 \nabla^2 \Phi_2 \Phi_3
\non \\
\non \\ 
(5l) &\qquad& h_4^{(=)} \, \thb^2 {\cal F}^2 \nabla^2 \Phi_i \Phi_i \Phib^i \Phib^i \qquad ; \qquad  
 h_4^{(\neq)} \, \thb^2 {\cal F}^2 \nabla^2 \Phi_i \Phi_j \Phib^i \Phib^j \quad  i < j 
\non \\
\non \\
(5m) &\qquad& h_5 \, \thb^2 {\cal F}^2 \Phi_i \Phib^i \Phib_1 \Phib_2 \Phib_3
\eea
We have not explicitly indicated background or quantum matter fields since it should be clear from the context. 
For instance, $\Phi_i \Phib^i$ stands for $\Phi_i \Phib^i_q$ or $(\Phi_i)_q \Phib^i$. 

We note that all vertices containing quantum gauge fields are at least of order $\Fab$. 
Hence vertices with quantum gauge fields and order ${\cal F}^2$ could be only employed in tadpole 
diagrams which vanish in dimensional regularization.
This is the reason why in vertices $(5d, 5e)$ we take gauge fields to be only background.
 
The expressions for the vertices of the one--flavor case can be obtained from the previous ones by 
dropping flavor indices and setting 
\bea
&& h_1 = -h_2 = h/2 \quad , \qquad  \hb_1 = -\hb_2 = \hb/2
\non \\
&& h^{(=)}_4 = h_4 \qquad \qquad , \qquad h_4^{(\neq)} = 0
\eea 
Moreover, we need take into account extra symmetry factors that
arise when specifying quantum or background matter. For instance, the term $\Phi^3$ in $(5f)$ would give rise
to $3 \Phi^2 \Phi_q$. The vertex $(5h)$ is absent for trivial symmetry reasons.

\begin{figure}
  \includegraphics[width=\textwidth]{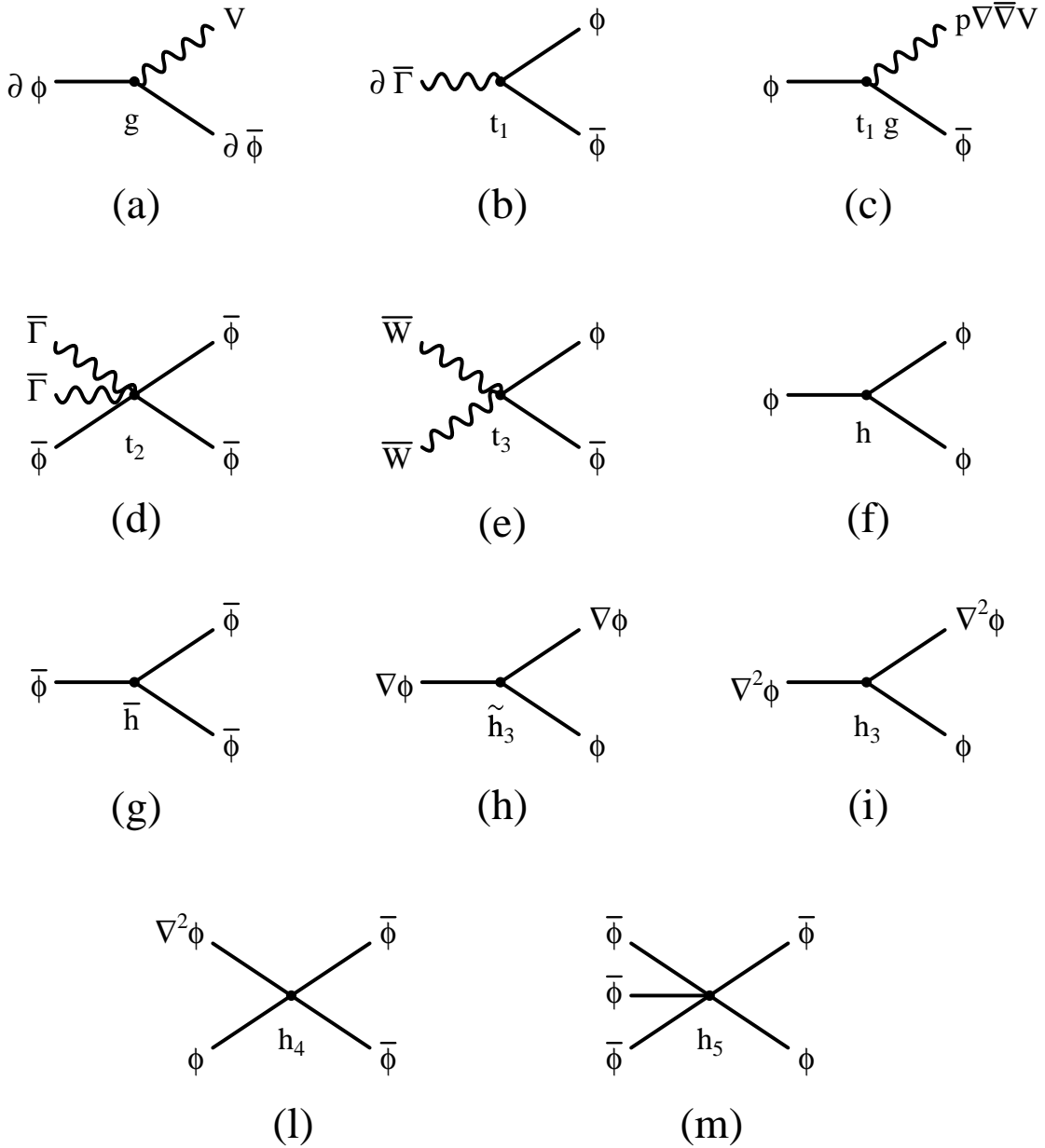} 
  \caption{Vertices from the actions (\ref{classaction1}, \ref{classaction2}).} 
  \label{fig:vertices}
\end{figure}

\newpage


\end{document}